%
%
%
%
%
%
%
%
%
%
%
%
%
\magnification=1100
\font\t=cmcsc10 at 13 pt
\font\tt=cmcsc10
\font\n=cmcsc10
\font\foot=cmr9
\font\foots=cmsl9
\font\abs=cmr8
\font\babs=cmbx8
 
\centerline{\t Notion of Parallelism on a Generic Manifold} 
\centerline{\t and Consequent Geometrical Specification} 
\centerline{\t of the Riemannian Curvature\footnote{\dag}
{\foot{Rend. Circ. Mat. Palermo, (1917), 
{\bf 42}, pp.~173-204.}}}\bigskip
\centerline{Memoir by}\bigskip
\centerline{\n Tullio LEVI-CIVITA}\bigskip
\vbox to 0.6 cm {}
\centerline{\tt translation and foreword by}\smallskip
\centerline{Marco Godina{\footnote{$^*$}{\foot SIGRAV, Viale F. Crispi 3 - 67100 L'Aquila (Italy).}} and Julian Delens}\medskip
{\babs Foreword.} {\abs With regard to classical differential geometry, this paper written in 1916 by T. Levi-Civita introduces the notion of parallelism for a Riemannian manifold of arbitrary dimensions. It also provides a geometrical explanation for the Riemannian curvature and at the same time significantly reduces the mathematical formalism compared with the scientific literature of the time. }
\bigskip
\centerline{Introduction}
\medskip
Einstein's theory of relativity (by now even backed by the explanation of the famous secular inequality, that can be detected in Mercury's perihelion and that Newton's law can't predict) considers the geometrical structure of space as being very weakly dependent, albeit fully, from the physical phenomena that take place within it. This differs from the classical theories that all assume the physical space as given a priori. Mathematically implementing Einstein's grand idea (which has in Ricci's absolute differential calculus its obvious algorithmic tool) involves the curvature of a certain four-dimensional manifold and the use of its corresponding Riemann symbols. Coming across, or rather having to continuously handle such symbols in regard to matters of such a high general interest has led me to investigate if it shall not be possible to significantly reduce the formalism that is usually needed to introduce them and to set their covariant behaviour \footnote{($^1$)}{\foot{Cf., e.g., L.~Bianchi,~}\foots{Lezioni di geometria differenziale,}
\foot{Vol. I (Pisa, Spoerri, 1902), pp. 69-72.~}}.
\par An improvement in this respect is actually possible and essentially makes up the content of paragraphs 15 and 16 of the present paper; which was initially conceived only with this goal but has gradually expanded in order to make proper room also for the geometrical interpretation.
\par At first I had thought I had definitively found it in  Riemann's original works ``{\it {\" U}ber die Hypothesen welche Geometrie zu Grunde liegen\/}'' and ``{\it Commentatio mathematica\/} \dots'' \footnote{($^2$)}{\foot{B.~Riemann,~}\foots{Gesammelte mathematische Werke,}
\foot{(Leipzig, Teubner, 1876), pp. 261-263, 381-382.~}}; but there is just a initial nucleus. On one hand in fact, once the mentioned sources have been put together, one has the impression that Riemann actually had in mind that very characterization of the intrinsic and invariant curvature, that will be explained here (paragraphs 17-18). 
\vfil
\eject
\noindent But on the other hand there is no trace, nor in  Riemann's paper, nor in the explanatory comment by Weber \footnote{($^3$)}{\foot{loc.~cit. ($^2$),~}\foot{pp. 384-389.~}}, of those clarifications (the parallel directions on a general manifold concept and taking into consideration a geodetic infinitesimal quadrangle with two parallel sides) that are to be understood as being essential from a geometrical point of view. Furthermore the justification for the formal step that according to Riemann should lead from the undisputable premises to the just as much undisputable final curvature expression can't be found---or at least I can't find any.
\par In a final critical note I will let the reader know about this doubt of mine and I will as well provide the elements needed to judge it.
\par The first and lengthiest part of this memoir (paragraphs 1-14) is focused on introducing and explaining the concept of parallelism in a manifold $V_n$  with any metric. 
\par So let's start from an infinitesimal portion of space and try to define the parallelism of two directions $(\alpha )$, $(\alpha')$ coming from two infinitely close points $P$ and $P'$. For this purpose let's recall that any manifold $V_n$ can be considered as immersed in an Euclidean space $S_N$ with a sufficiently large number $N$ of dimensions, and first of all notice that, considering that a generic direction $(f)$ of $S_N$  shall be drawn from point $P$, the ordinary parallelism in such a space, would imply:
$$
{\rm angle}\,\widehat{(f)(\alpha)} = {\rm angle}\,\widehat{(f)(\alpha')}\, ,
$$
\noindent for any $(f)$. In such a way, the parallelism in $V_n$ will be defined, as long as we require that this condition shall be satisfied for all the $(f)$ directions belonging to $V_n$ (i.e., to the set of directions of $S_N$  tangent to $V_n$ in $P$). 
\par To back this definition we have to note that, while it reproduces, as required, the elementary behavior of the Euclidean manifolds $V_n$, it anyhow has an intrinsic nature, because it essentially depends only on the metric of $V_n$ and not also on the auxiliary ambient space $S_N$. In fact our definition of parallelism has the following analytical translation: Once $V_n$ is referred to the $x_i\,$ $(i=1 ,2,\dots,n)$ general coordinates, shall the $dx^i$ be the increments that correspond to passing from $P$ to $P'$; $\xi^{(i)}$ the parameters of a generic direction $(\alpha)$ coming from $P$; $\xi^{(i)} +d\xi^{(i)}$  those belonging to an infinitely close direction $(\alpha')$ and drawn from the point $P'$. The condition of parallelism is given by the $n$ equations
$$
d\xi^{(i)} + \sum\limits_{j,l=1}^n\Bigl\{ ^{jl}_{\, i} \Bigr\} \, dx_j \, \xi^{(l)}   = 0 \qquad\qquad\qquad (i=1 ,2,\dots,n),
\leqno(A)
$$
\noindent where $\Bigl\{ ^{jl}_{\, i} \Bigr\} $ are the well-known Christoffel's symbols.
\par Once the law that allows passing from a point to an infinitely close point is secured, we are definitively able to carry out the transport of parallel directions along any curve $C$. If $x_i = x_i(s)$ are its parametric equations, 
\vfil
\eject
\noindent it's obviously sufficient to consider, in equations $(A)$, the coordinates $x_i$ and subordinately the $\Bigl\{ ^{jl}_{\, i} \Bigr\} $ as assigned functions, the $\xi^{(i)}$ as functions to be determined of the parameter $s$, and one has the ordinary linear system 
$$
{d\xi^{(i)}\over {ds}} + \sum\limits_{j,l=1}^n\Bigl\{ ^{jl}_{\, i} \Bigr\}  \, {dx_j\over {ds}} \, \xi^{(l)}   = 0 \qquad\qquad\qquad (i=1 ,2,\dots,n),
$$
\noindent reducible to a typical form (called a skew determinant), which already
appeared in other researches and was the object of a systematic investigation by Mr.~Eiesland  \footnote{($^4$)}{\foot{J.~Eiesland ,~}\foots{On the Integration of a System of Differential Equations in Kinematics~}\foot{[American Journal of Mathematics, vol. XXVIII (1906), pp. 17-42].}}, Laura \footnote{($^5$)}{\foot{E.~Laura,~}\foots{Sulla integrazione di un sistema di quattro equazioni differenziali lineari a determinante gobbo per mezzo di due equazioni di Riccati~}\foot{[Atti della R. Accademia delle Scienze di Torino, vol. XLII, (1906-1907), pp. 1089-1108; vol. XLIII, (1907-1908), pp. 358-378].}}, Darboux \footnote{($^6$)}{\foot{G.~Darboux,~}\foots{Sur certains syst{\` e}mes d'{\' e}quations lin{\' e}aires~}\foot{[Comptes rendus hebdomadaires des s{\' e}ances de l'Acad{\' e}mie des Sciences, t. CXLVIII (1er semestre 1909), pp. 16-22], and \foots{Sur les syst{\` e}mes d'{\' e}quations diff{\' e}rentielles homog{\` e}nes} \foot{(Ibid., pp. 673-679 and pp. 745-754).}}}, Vessiot \footnote{($^7$)}{\foot{E.~Vessiot,~}\foots{Sur l'int{\' e}gration des syst{\` e}mes lin{\' e}aires {\` a} d{\' e}terminant gauche ~}\foot{[Comptes rendus hebdomadaires des s{\' e}ances de l'Acad{\' e}mie des Sciences, t. CXLVIII (1er semestre 1909), pp. 332-335].}}.
\par Here are some geometrical outcomes.
\par $1^{\rm st}$ The parallel direction in a generic point $P$ to a direction $(\alpha)$ coming from any another point $P_{0}$ generally depends on the path taken when passing from $P_{0}$  to $P$. The independence of this path is an exclusive property of Euclidean manifolds.
\par $2^{\rm nd}$ Along the same geodesic, the directions of the tangents are parallel, this generalises an obvious characteristic of the Euclidean spaces (precisely the one Euclid places as first of the elements as an early idea of straight line).  
\par $3^{\rm rd}$ The transport by parallelism, along any path, of two concurrent directions preserves their angle. What is meant by this is that obviously the angle given by two generic directions coming from a same point is also the angle by their parallels in any other (given) point. Keeping into consideration these geodesic properties that have been found, the corollary that can be drawn is that, along a geodesic, parallel directions are always equally inclined on the geodesic itself. If specifically a $V_2$ is considered, this is also a sufficient condition; therefore, for ordinary surfaces, the parallelism along a geodesic also means isogonality.
\par I haven't shown the content of the various paragraphs in an orderly fashion. A look at the summary at the end of this paper will easily remedy this.
\bigskip
\centerline{1. {\bf Preliminaries}}
\medskip
Let (with the usual symbols)
$$
ds^{2} = \sum\limits_{i,k=1}^n a_{ik} \, dx_{i} \, dx_{k}
$$
\vfil
\eject
\noindent be the expression for the square of the line element of any manifold $V_n$.
\par As is well known, $V_n$ can always be considered as being immersed in an Euclidean space $S_N$ having a number of dimensions which is sufficiently big (but no greater than $n(n+1)/2$).
\par \noindent Let $y_{\nu}$ $(\nu=1 ,2,\dots,N)$ be the Cartesian coordinates in such space. Within which, it will be helpful to consider $V_n$ as defined by the parametric expressions
$$
y_{\nu} = y_{\nu}(x_1 , x_2,\dots, x_n) \qquad\qquad\qquad (\nu=1 , 2,\dots, N)
 \leqno(1)
$$
\noindent of the Cartesian coordinates in terms of the intrinsic coordinates, accordingly we have
$$
ds^{2} = \sum\limits_{\nu=1}^N dy^2_{\nu} = \sum\limits_{i,k=1}^n a_{ik} \, dx_{i} \, dx_{k}\, .
\leqno(2)
$$
\par Given any curve $C$ within $V_n$, and setting
$$
x_i = x_i(s)  \qquad\qquad\qquad (i=1 ,2,\dots,n)
\leqno(3)
$$
\noindent its parametric equations, where $s$ denotes the arc length (measured from an arbitrary origin). Obviously $C$ also belongs to the ambient space $S_N$, and as such it is defined by the parametric expressions $y_i(s)$ of its  Cartesian coordinates. These parametric expressions are definitively given by (1) for which the values $x_i$ are provided by (3). Differentiating them with respect to $s$, and marking the derivatives with respect to this argument with prime marks, we get
$$
y'_{\nu} = \sum\limits_{i=1}^n{ \partial y_{\nu}\over {\partial x_i}}x'_i\qquad\qquad\qquad (\nu=1 , 2,\dots, N)\, . \leqno(4)
$$
\par Let us refer to a generic, but well specified, value for $s$, i.e., any point $P$ belonging to the $C$ curve. The $y'_{\nu}$ clearly are the direction cosines for $C$ in $P$ with regard to the coordinate axis of the Euclidean space $S_N$; the $x'_{\nu}$ are the {\it direction parameters\/} (of $C$ itself as well as in the same point $P$) in regards to $V_n$.
\par  Bear in mind \footnote{($^8$)}{\foot{L.~Bianchi, loc.~cit. ($^1$),~}\foot{pp. 365-367.~}} that, if we set
$$
y''_{\nu} = c \, q_{\nu} \, , 
\leqno(5)
$$
\noindent with $c \ge 0$ and $\sum\limits_{\nu=1}^N q^2_{\nu} =1$, the curvature $c$ of $C$ will be defined in $P$, as well as (excluding the exception $c = 0$),  by means of the cosines $q_{\nu}$, a direction $(q)$, called the {\it absolute principal normal\/} at the point $P$. Projecting it (orthogonally) onto the hyperplane which is tangent to $V_n$ in $P$ we will find a direction $(q^{\ast})$, which is also normal to the curve, called the {\it relative principal normal\/}.
\par We shall denote by $q^{\ast}_{\nu}$ the  direction cosines of $(q^{\ast})$; by $\Phi$ the angle between $(q)$ and $(q^{\ast})$; and finally by $\alpha_{\nu}$ the direction cosines of a generic direction $(\alpha)$ coming from $P$ and belonging
\vfil
\eject
\noindent to $V_n$ (i.e., to its tangent hyperplane). Clearly it will be
$$
\sum\limits_{\nu=1}^N \alpha_{\nu} \, q_{\nu} = \cos \Phi \, \sum\limits_{\nu=1}^N \alpha_{\nu} \, q^{\ast}_{\nu}\, ,
$$
\noindent then, multiplying by $c$ and considering equations (5),
$$
\sum\limits_{\nu=1}^N \alpha_{\nu} \, y''_{\nu} = c \, \cos \Phi \, \sum\limits_{\nu=1}^N \alpha_{\nu} \, q^{\ast}_{\nu}\, .
$$
\par\noindent It is worth noting that the character of the second side is intrinsic compared with the manifold $V_n$, meaning that it can be interpreted independently from the Euclidean ambient space. In fact
$$
c \, \cos \Phi = \gamma
$$
\noindent is nothing other than the geodesic curvature of $C$, and $\sum\limits_{\nu=1}^N \alpha_{\nu} \, q^{\ast}_{\nu}$ is the cosine of the angle $\chi$  between $(\alpha)$ and the relative principal normal $(q^{\ast})$, such directions both belonging to $V_n$. Therefore we have the equality 
$$
\sum\limits_{\nu=1}^N \alpha_{\nu} \, y''_{\nu} = \gamma \, \cos\chi
\leqno(6)
$$
\noindent (depending on the given curve $C$) for any direction $(\alpha)$ of $V_n$.
\bigskip
\centerline{2. {\bf Parallel directions in $V_n$ along a predetermined curve}}
\medskip
Let's suppose that for every point $P$ belonging to $C$ there corresponds a direction $(\alpha)$ belonging to $V_n$. Therefore the direction parameters $\xi^{(i)}$ $(i=1 ,2,\dots,n)$  that define $(\alpha)$ within $V_n$, as well as the direction cosines $\alpha_{\nu}$ $(\nu=1 , 2,\dots, N)$ that identify it in the ambient space, have to be considered as functions of $s$; we then get [as previously for the direction of $C$, in accord with equations (4)]
$$
\alpha_{\nu} = \sum\limits_{l=1}^n{ \partial y_{\nu}\over {\partial x_l}}\xi^{(l)}\, .
\leqno(7)
$$
\par\noindent Let's imagine to vary $P$ along $C$. The condition of ordinary parallelism for directions $(\alpha)$ (with respect to the ambient space $S_N$) implies that the angles that they form with any same direction that is chosen shall be equal.
\par In order to achieve a concept of parallelism which relates only to $V_n$ , let us consider the elementary  phenomenon, i.e., passing from $P$ to an infinitely close point.
\par Setting $(f )$ as a generic {\it fixed\/} direction of $S_N$, where $f_{\nu}$ are the corresponding direction cosines. When we increment $s$ by $ds$, the cosine of the angle between $(\alpha)$ and $(f )$, 
$$
\sum\limits_{\nu=1}^N \alpha_{\nu} \, f_{\nu}\, ,
$$
\noindent is increased by 
$$
ds \sum\limits_{\nu=1}^N \alpha'_{\nu} \, f_{\nu}\, .
$$
\vfil
\eject
\noindent The ordinary parallelism would imply that such increase shall be reduced to zero for all the $(f )$ directions, and would implies $\alpha_{\nu}$ is constant.
\par {\it Let's be satisfied in requiring that the angle between $(\alpha)$ and any $(f)$ shall remain unchanged for the directions that belong to $V_n$\/}, i.e., that the $ds \sum\limits_{\nu=1}^N \alpha'_{\nu} \, f_{\nu}$ increase shall become zero [not for all the $(f)$, but only] for these tangential directions.
\par If such directions are seen to be all and only those that are compatible with the constraints (1), then it's manifest (when the $f_{\nu}$ direction cosines are replaced with proportional quantities) that the condition set out is equivalent to the following:  
$$
\sum\limits_{\nu=1}^N \alpha'_{\nu} \, {\delta y}_{\nu} = 0
\leqno(I)
$$
\noindent {\it for all the ${\delta y}_{\nu} $ displacements that can be compatible with the (1) costraints\/}.
\par As we have from equation (1)
$$
{\delta y}_{\nu} = \sum\limits_{k=1}^n{ \partial y_{\nu}\over {\partial x_k}}\, {\delta x}_k
$$
\noindent where the $\delta x_k$ are entirely arbitrary, the (I) is split in the $n$ equations
$$
\sum\limits_{\nu=1}^N {\alpha}'_{\nu} \, { \partial y_{\nu}\over {\partial x_k}} = 0\qquad\qquad\qquad (k=1 , 2,\dots, n) , \leqno(8)
$$
\noindent which are a formal translation of the parallelism for the $(\alpha)$ directions along $C$.
\bigskip
\centerline{3. {\bf Intrinsic form for the parallelism conditions}}
\medskip
The $\alpha_{\nu}$  and $y_{\nu}$ entities in equations (8) relate to the ambient space. They can be removed substantially by involving only elements that belong to the manifold $V_n$.
\par Specifically let's start to replace the $\alpha_{\nu}$ direction cosines with their expressions (7) as a function of the direction parameters $\xi^{(l)}$. Differentiating them with respect to the arc length $s $ of $C$, we get
$$
{\alpha}'_{\nu} = \sum\limits_{l=1}^n { \partial y_{\nu}\over {\partial x_l}} \, { d \xi^{(l)}\over {ds}}+
\sum\limits_{j,l=1}^n { \partial^2 y_{\nu}\over {\partial x_j\partial x_l}} \,  x'_{j} \, \xi^{(l)}\, .
$$
\par\noindent On the other hand under (2) we have
$$
a_{kl}=\sum\limits_{\nu=1}^N { \partial y_{\nu}\over {\partial x_k}}\, { \partial y_{\nu}\over {\partial x_l}} \qquad\qquad\qquad (k, l=1 , 2,\dots, n) ,  
$$

\vfil
\eject
\noindent and this leads, thanks to the Christoffel symbols of the first kind, to
$$
\eqalignno{
a_{jl,k} &= {1 \over 2}\biggl( {{\partial a_{kl}} \over {\partial x_{j}}} + {{\partial a_{jk}} \over {\partial x_{l}}}  - {{\partial a_{jl}} \over {\partial x_{k}}}  \biggr) \cr
          &= {1 \over 2} \sum\limits_{\nu=1}^N \biggl[  {{\partial} \over {\partial x_j}}\biggl({{\partial y_{\nu}} \over {\partial x_k}}\, {{\partial y_{\nu}} \over {\partial x_l}}\biggr) + {{\partial} \over {\partial x_l}}\biggl({{\partial y_{\nu}} \over {\partial x_j}}\, {{\partial y_{\nu}} \over {\partial x_k}}\biggr) - {{\partial} \over {\partial x_k}}\biggl({{\partial y_{\nu}} \over {\partial x_j}}\, {{\partial y_{\nu}} \over {\partial x_l}}\biggr)  \biggr] \cr
          &=\sum\limits_{\nu=1}^N  {{\partial^2 y_{\nu}} \over {\partial x_j \partial x_l}} \, {{\partial y_{\nu}} \over {\partial x_k}}& (i, l, k=1 , 2,\dots, n). 
}
$$
\noindent So the (8) become
$$
\eqalignno{
\sum\limits_{l=1}^n a_{kl} \, {d\xi^{(l)}\over {ds}} + \sum\limits_{j,l=1}^n a_{jl,k}  \, x'_{j} \, \xi^{(l)}   &= 0 &(k=1 ,2,\dots,n).
}
$$
\noindent When these are multiplied by $a^{(ik)}$ \footnote{($^9$)}{\foot{The coefficients $a^{(ik)}$ are the components of the reciprocal form associated to the fundamental form $ds^{2} = \sum\limits_{i,k=1}^n a_{ik} \, dx_{i} \, dx_{k}\, .$}}, and are summed over $k$ (from $1$ to $n$), and bearing in mind the following definitions
$$
\eqalignno{
\Bigl\{ ^{jl}_{\, i} \Bigr\}&= \sum\limits_{k=1}^n a_{jl,k}\, a^{(ik)} & (j, l, i=1 ,2,\dots,n)
}
$$
\noindent for the Christoffel symbols of the second kind, then we get the following equivalent equations
$$
{d\xi^{(i)}\over {ds}} + \sum\limits_{j,l=1}^n\Bigl\{ ^{jl}_{\, i} \Bigr\} \, x'_{j}  \, \xi^{(l)}   = 0 \qquad\qquad\qquad (i=1 ,2,\dots,n),
\leqno(I_a)
$$
\noindent {\it that define how the  $\xi^{(i)}$ parameters vary along  $C$, based on the condition that the directions that these parameters identify are kept parallel.\/}
\par Since $C$ is considered as assigned a priori  (and together with the $s$-function  $x$ and $x'$ expressions), the  $\sum\limits_{j=1}^n\Bigl\{ ^{jl}_{\, i} \Bigr\} \, x'_{j} $ coefficients for each  $\xi^{(i)}$ in the  $(I_a)$ functions should be seen as known functions of the independent variable  $s$. The $(I_a)$ themselves are coherent as  $n$ ordinary differential equations in as many $\xi^{(i)}$ quantities. According to the well-known existence theorems, this yields that when any direction is drawn from any point $P_{0}$ belonging to  $C$, the directions that are parallel through any other point $P$ belonging to the curve will be determined.
\bigskip
\centerline{4. {\bf Comparison with the Euclidean behaviour.}}
\centerline{{\bf Its characteristic property in respect of parallelism}}
\medskip
Based on what we have just seen, the elementary property, for which from a point $P$ just one direction that is parallel to any other assigned direction can exit through $P_{0}$, continues to exist along the curve $C$.
\vfil
\eject
\noindent Anyhow, it's worth noting that, while for the Euclidean spaces the parallel through $P$ is unique in an absolute sense, within our $V_n$ whose metric is arbitrary, it will generally depend on $C$, i.e., on the path along which we can pass from $P_{0}$ to $P$.
\par We can indeed add that, {\it if\/}(through any point $P$ belonging to a certain portion of space) {\it the parallel\/} (to any direction drawn trough another point $P_{0}$ belonging to that portion of space) {\it is independent from the path, then the space\/} $V_n$ (in that portion of space) is {\it necessarily Euclidean\/}.
\par In fact, note that from the $(I_a)$, multiplying by $ds$ and setting for the sake of brevity
$$
X^{(i)}_j = \sum\limits_{l=1}^n\Bigl\{ ^{jl}_{\, i} \Bigr\} \, \xi^{(l)}  \, ,
$$
\noindent yields
$$
d\xi^{(i)} = - \sum\limits_{j=1}^n X^{(i)}_j \, dx_j  \, .
\leqno(9)
$$
\par The $d\xi^{(i)}$, and therefore the right hand side of (9), must be exact differentials if the path has to be independent as required. When a second system of $\delta x_j$ increments of $x_j$, that are independent from the $dx_j$ and such that $d\delta x_j = \delta dx_j$, is designated this translates into the $n$  identities
$$
\eqalignno{
\delta \, \sum\limits_{j=1}^n X^{(i)}_j \, dx_j  = d \, \sum\limits_{j=1}^n X^{(i)}_j \, \delta x_j    &\ &(i=1 ,2,\dots,n).
}
$$
\par Making this explicit and noting that the equality must exist, whatever the two systems of increments $dx_j$, $\delta x_j$ may be, as well as the $\xi^{(l)}$ values (initial, and therefore also generic), we are automatically led to show that all the Riemann symbols will be reduced to zero \footnote{($^{10}$)}{\foot{From a methodological point of view, if the ordinary use of Riemann's symbols and curvatures had  to actually be considered preferable to the one that will be described in paragraphs 15-19, then the theorem of the text should appear after those paragraphs. I am bringing this forward for the ease of the readers who are familiar with Riemann's symbols.}}, i.e., that this is a Euclidean manifold.\qquad {Q.E.D.}
\bigskip
\centerline{5. {\bf An other form for the $(I_a)$ equations --- Single function dependence}}
\medskip
In the $(I_a)$ equations the direction parameters $\xi^{(i)}$ ($= {{dx_i}\over {ds}}$, as $ds$ represents the length of an element drawn along the α direction $(\alpha)$, and $dx_i$ the corresponding $x_i$ coordinate increase) are the elements that determine the parallel directions $(\alpha)$. These $\xi^{(i)}$ form a contravariant system   \footnote{($^{11}$)}{\foot{Cf. (at this point only for the expressions) G.~Ricci et T.~Levi-Civita ,~}\foots{M{\` e}thodes de calcul diff{\' e}rentiel absolu et leurs applications~}\foot{[Mathematische Annalen, Bd. LIV (1900), pp. 125-201].}} (with respect to any coordinates' transformation $x_i$). 
\vfil
\eject
\noindent In some cases it is preferable to use the related covariant system
$$
\xi_{i} = \sum\limits_{k=1}^n a_{ik}\, \xi^{(k)}  \, ,
\leqno(10)
$$
\noindent i.e., the so called {\it moments\/}, instead of the contravariant system $\xi^{(i)}$.
\par In order to get the covariant form for the $(I_a)$, we shall differentiate equations (10) by replacing the ${{d \xi^{(k)}}\over {ds}}$ with the expressions provided by the $(I_a)$. We will get (once we have inverted the index of summation $k$ into $l$ in the last term)
$$
{d\xi_{i}\over {ds}} = - \sum\limits_{k,j,l=1}^{n} a_{ik}\, \Bigl\{ ^{jl}_{\, k} \Bigr\} \, x'_{j}  \, \xi^{(l)} + \sum\limits_{l=1}^{n} {{d a_{il}}\over {ds} }\, \xi^{(l)} \, .
$$
\par We have the identity 
$$
\sum\limits_{k=1}^{n} a_{ik}\, \Bigl\{ ^{jl}_{\, k} \Bigr\} = a_{jl,i}\, ,
$$
\noindent and (also is)
$$
{{da_{il}}\over {ds}} = \sum\limits_{j=1}^{n} {{\partial a_{il}}\over {\partial x_j}}\, x'_{j}  = \sum\limits_{j=1}^{n} \bigl(a_{jl,i} + a_{ij,l}\bigr) \, x'_{j}\, ,
$$
\noindent it follows
$$
{d\xi_{i}\over {ds}} = \sum\limits_{j,l=1}^{n} a_{ij,l}\, x'_{j} \, \xi^{(l)} \, .
\leqno(11)
$$
\par The moments (not the $\xi^{(l)}$) must also appear on the right hand side: this is quite easy in accord with (10). These when they are solved provide the
$$
\xi^{(l)} = \sum\limits_{k=1}^{n} a^{(lk)}\xi_k \, .
$$
\par Thus, considering the expressions for the symbols of Christoffel of the second kind that were mentioned in the previous paragraph, it gives
$$
\sum\limits_{j,l=1}^{n} a_{ij,l}\, x'_{j} \, \xi^{(l)}  =  \sum\limits_{j,k=1}^{n} \Bigl\{ ^{ij}_{\, k} \Bigr\} \, x'_{j}  \, \xi_{k}   \, ,
$$
\noindent it follows that (replacing $l$ for $k$ as index of summation) the equations transformed into
$$
{d\xi_{i}\over {ds}} = \sum\limits_{j,l=1}^n\Bigl\{ ^{ij}_{\, l} \Bigr\} \, x'_{j}  \, \xi_{l}  \qquad\qquad\qquad (i=1 ,2,\dots,n).
\leqno(I_b)
$$
\par One can see that such $(I_b)$ is the adjoint system of the previous $(I_a)$. In fact, the coefficient of $ \xi_{l}$ in the $i^{\rm th}$ equation $(I_b)$ is equal and opposite to the coefficient of $\xi^{(i)}$ in the $l^{\rm th}$ equation  $(I_a)$.
\par A third aspect can be attributed to these equations. Although actually less descriptive, it recalls the classical Lagrangian equations of dynamics.
\par The analogy lies in what leads to a single function with $3n$ variables $x_i $, $x'_{i}$, $\xi^{(i)}$ ,
$$
B = \sum\limits_{i,j=1}^n a_{ij} \, x'_{i}  \, \xi^{(j)} \, ,
\leqno(12)
$$
\vfil
\eject
\noindent that is bilinear in the  $\xi^{(j)}$ (that act as unknowns) and in the  $x'_{i}$ (which are, just as the $x_i $, the assigned functions of $s$).
\par We get from (10)
$$
\xi_i = {{\partial B}\over {\partial x'_{i}}} \, ,
$$
\noindent and it's immediately evident (as the indices have been obviously interchanged) since
$$
a_{ij,l} = {1\over 2}\biggl( {{\partial a_{lj}}\over {\partial x_i}} + {{\partial a_{il}}\over {\partial x_j}} - {{\partial a_{ij}}\over {\partial x_l}}\biggr)
$$
\noindent that the right hand side of (11) can be written as
$$
{1\over 2}{{\partial B}\over {\partial x'_{i}}} + {1\over 2} \sum\limits_{j=1}^n{{\partial}\over {\partial x_{j}}}\biggl({{\partial B}\over {\partial x'_{i}}}\,  x'_{j} - {{\partial B}\over {\partial \xi^{(i)}}} \, \xi^{(j)}\biggr)\, .
$$
\par Therefore, from (11) (which is essentially equivalent to both $(I_a)$ and $(I_b)$) we have the identity 
which only depends on the function  $B$:
$$
{d\over {ds}} {{\partial B}\over {\partial x'_{i}}} = {1\over 2}{{\partial B}\over {\partial x_{i}}} + {1\over 2} \sum\limits_{j=1}^n{{\partial}\over {\partial x_{j}}}\biggl({{\partial B}\over {\partial x'_{i}}}\,  x'_{j} - {{\partial B}\over {\partial \xi^{(i)}}} \, \xi^{(j)}\biggr)\, .
\leqno(I_c)
$$
\bigskip
\centerline{6. {\bf Quadratic integral --- Angle preservation.}}
\centerline{{\bf Composition of orthogonal solutions}}
\medskip
The $(I_a)$ linear equations imply the quadratic integral
$$
\sum\limits_{i,j=1}^n a_{ij} \, \xi^{(i)}   \, \xi^{(j)}  = {\rm const.}
$$
\par The most immediate way to see this is by noting that, thanks to the (10), the left hand side can be written as
$$
\sum\limits_{i=1}^n \xi^{(i)}   \, \xi_i \, ,
$$
\noindent and therefore its derivative is
$$
\sum\limits_{i=1}^n \biggl( { {d\xi^{(i)}} \over {ds}} \, \xi_i \, +\, { {d\xi_i} \over {ds}}  \, \xi^{(i)} \biggr)\, ,
$$
\noindent that in the same way is reduced to zero thanks to the $(I_a)$ and to the equivalent $(I_b)$.
\par This means that, even without any direct verification, we can state that $\sum\limits_{i=1}^n \xi^{(i)}   \, \xi_i$ is constant, according to a well-known property of the adjoint systems. In fact (previous paragraph),  $\xi^{(i)} $ and $\xi_i$ are solutions of these two systems.
\par The obvious corollary of the existence of the quadratic integral is that, if the initial values of
\vfil
\eject
\noindent the $\xi^{(i)} $ are really direction parameters, and, as such, provide
$$
\sum\limits_{i,j=1}^n a_{ij} \, \xi^{(i)}   \, \xi^{(j)}  = 1\, ,
\leqno(13)
$$
\noindent this equation stands for any  $s$, that is, the  $\xi^{(i)}$ solutions preserve the character of being direction parameters along $C$: such circumstance is implicit in how the problem has been set, but its analytic expression by means of the $(I_a )$ does not make this obvious a priori.
\par Another noteworthy corollary concerns the preservation along $C$ of the angle $\vartheta$ of two directions that are transported by parallelism. Let $\xi^{(i)} $ and $\eta^{(i)}$ be the parameters of these two directions and  $\xi_i $, $\eta_i$ the corresponding moments. We get
$$
\cos \vartheta = \sum\limits_{i,j=1}^n a_{ij} \, \xi^{(i)}   \, \eta^{(j)}  = \sum\limits_{i=1}^n \xi^{(i)} \, \eta_i  = \sum\limits_{i=1}^n \xi_i \, \eta^{(i)} \, .
$$
\medskip For example, let us refer to the last expression, and let's differentiate having replaced the $ {d\xi_i} \over {ds}$ with the values that are provided by the $(I_b )$ and the $ {d\eta^{(i)}} \over {ds}$ with those that are provided by the $(I_a )$ (having changed the $\xi$ into $\eta$). Here too the result is reduced to zero.\qquad {Q.E.D.}
\par If for example we refer to the $(I_a )$, the linearity of the parallelism equations imply that, if the $\xi^{(i)}$ and $\eta^{(i)}$ are solutions, then also any one of their linear combinations with constant coefficients shall be. Let's suppose that specifically $\xi^{(i)} $ and $\eta^{(i)}$ shall be the parameters of two orthogonal directions (as well as being parallel) and let's assume
$$
\zeta^{(i)}  = \cos \vartheta \, \xi^{(i)}  + \sin \vartheta \, \eta^{(i)} 
$$
\noindent where $\vartheta$ constant. We obviously get
$$
\sum\limits_{i,j=1}^n a_{ij} \, \zeta^{(i)}   \, \zeta^{(j)}  = 1\, ;\qquad \sum\limits_{i=1}^n \zeta^{(i)} \, \xi_i =\cos \vartheta \, , \qquad \sum\limits_{i=1}^n \zeta^{(i)} \, \eta_i = \sin \vartheta \, ,
$$
\noindent therefore the $\zeta^{(i)}$ are the parameters of a direction that not only satisfies the condition of parallelisms along $C$ but also belongs to the plane that is identified by the first two and in every point it always forms the same angles $\vartheta$ and ${\pi \over 2} - \vartheta$ with them.
\par This finding can easily be extended to any required mutually orthogonal solution and provides the following descriptive
proposition. {\it Every direction that remains rigidly connected with the parallel directions while it moves along $C$ shall also satisfy the condition of parallelism\/}.
\bigskip
\centerline{7. {\bf Geodesics --- Their characteristic property in respect of the direction}}
\medskip
The equations for the geodesics of  $V_n$ are known to be
$$
x''_i + \sum\limits_{j,l=1}^n\Bigl\{ ^{jl}_{\, i} \Bigr\} \, x'_j  \, x'_l   = 0 \qquad\qquad\qquad (i=1 ,2,\dots,n).
$$
\vfil
\eject
These imply that if $C$ is a geodesic, the $(I_a )$ allow the solution $\xi^{(i)}=x'_i $, and vice-versa. Therefore, {\it the direction of a geodesic in any of its points is always parallel to the initial direction. Mutually any curve that has this property is a geodesic\/}.
\par So a discriminant of the straight lines within the Euclidean spaces is extended to the geodesics of a generic (Riemannian) manifold and at the same time it is established that the conservation of the direction and (the other characteristic of the geodesics i.e.) the property of minimizing the distance necessarily coincide.
\bigskip
\centerline{8. {\bf The inclination on the transport curve}}
\medskip
\par
It is worthwhile noting an expression for linear combinations of the $(I_a)$. In order to avoid the expansions, we shall trace back to formula  $(I)$ that is the root of these differential equations and that is essentially their equivalent. Such formula must hold for every displacement $\delta y_{\nu}$, and therefore also for every direction, that belongs to $V_n $. Specifically, let's consider the tangent direction to $C$, setting $y'_{\nu}$ into the $(I)$ in place of $\delta y_{\nu}$. It follows that
$$
\sum\limits_{\nu=1}^N \alpha'_{\nu} \, y'_{\nu} = 0\, ,
$$
\noindent that can be written as
$$
{d \over {ds}}\,\sum\limits_{\nu=1}^N \alpha_{\nu} \, y'_{\nu} = \sum\limits_{\nu=1}^N \alpha_{\nu} \, y''_{\nu} \, .
$$
Let us call $\psi$ the angle that the direction  $(\alpha)$ (whose parallelism we are considering) forms with $C$ and considering $(6)$, we get the equation
$$
{d \over {ds}}\, \cos \psi = \gamma \, \cos \chi \, .
\leqno(14)
$$
\noindent Bearing in mind (paragraph 1) that $\gamma$ represents the geodesic curvature of $C$, and $\chi$ the angle between $(\alpha)$ and the principal normal of $C$ (relative to the $V_n$). Then (14) provides the law with which the angle $\psi$ (on the same $C$) of a set of parallel directions varies along the transport curve $C$.
\par If $C$ is a geodesic ($\gamma=0$) and therefore $\cos \psi={\rm cost.}$, that is: {\it the parallel directions along a geodesic always form the same angle with this geodesic\/}. After all, this is a special case of angle conservation that has been seen in paragraph 6: it's sufficient to keep in mind (previous paragraph) that the directions of a geodesic are all parallel to each other.
\par As it is a Euclidean space, if $C$ is any curve then the $(14)$ identifies with the first Set of the Frenet formulae.
\bigskip
\centerline{9. {\bf The case of the ordinary space surfaces}}
\medskip
\noindent For $n=2$, the various directions coming from the same point of $C$ remain uniquely
\vfil
\eject
\noindent determined by the angle $\psi$, as long as qualitative information shall be linked to it: the direction that $\psi$ has to be measured from, starting from the $C$ direction. As a result (with an appropriate qualitative specification) equation (14) is able to uniquely define the parallel directions along a generic curve on a surface. The following can be specified: let $q^{\ast}$ be the relative principal normal (the projection of $C$'s principal normal on the surface's tangent plane), let's suppose to measure $\psi$ towards $q^{\ast}$  [meaning that it's measured starting from $C$'s positive direction, within the right angle that it forms with $q^{\ast}$].
\par 
\noindent If the angle $\psi$ is considered in this way it may not be the minimum angle between $C$ and $(\alpha)$ that formula (14) refers to (but the obtuse one that is its complementary); however, its cosine always coincides with the $\cos \psi$ of formula (14), being also (for the present $\psi$) $\sin \psi=\cos\chi$.
\par It follows that formula (14) takes the following simplified form
$$
{{d\psi} \over {ds}} = - \gamma\, .
\leqno(15)
$$
\noindent  In the previous section we already dealt with the case $\gamma = 0$ for any $n$. Regarding the surfaces, it may be added that the classical notion of geodesic parallelism is part of our general notion of parallelism. In fact, geodesically parallel curves admit geodesic curves as orthogonal trajectories; therefore they can be considered as parallel (according to the sense we have given to this definition) in relation to each of these geodesics since for both curves the following holds: $\psi ={ \pi \over 2}$, or $\psi ={ {3\pi }\over 2}$.
\par Let's set aside the case of $C$ being geodetic and let's consider a special example. Let's suppose we have a spherical surface, and that $C$ is a parallel. The direction $q^{\ast}$ at a generic point is the one of the meridian, towards $C$'s pole. If $\lambda$ is the latitude (that refers to the hemisphere that contains the $C$'s pole) and $R$ the radius of the sphere, then
$$
\gamma = {1\over {R\cos \lambda}} \, .
$$
\noindent On the other hand, if we imagine that we shall follow the parallel along the direction the longitude $\varphi$ grows,
$$
ds = R\cos \lambda \, d \varphi \, ,
$$
\noindent and then (15) is reduced to
$$
d\psi = - d \varphi\, .
$$
\noindent Therefore, the angle $\psi$ varies uniformly with longitude, and decreases by $2\pi$ within a full cycle. The direction that is parallel to one that is predetermined from the start turns out to be a {\it uniform\/} function of the points belonging to a parallel.
\par However, such uniformity does not necessarily stand for other closed curves: e.g., when the perimeter of a geodetic triangle is followed, $\psi$ is subject to a $2\pi - \varepsilon$ increment (where $\varepsilon$ represents the spherical excess). This can be recognized immediately considering that $\psi$ remains
\vfil
\eject
\noindent constant along the sides, and at the vertices it has sudden increases consisting in the supplementary angles.
\bigskip
\centerline{10. {\bf Spaces of constant curvature --- A remark on Clifford parallelism.}}
\medskip
We want to show how, for spaces of constant curvature \footnote{($^{12}$)}{\foot{Also in this regard the suggestion given in note ($^{10}$) has to be applied.~}} (in any dimension), the equations for parallelism {\it along a geodesic\/} are readily integrated, just as for the case that was examined in the previous paragraph, but in just two dimensions.
\par Just to focus, let's consider, without affecting generality, a $V_n$ manifold with a constant curvature = 1 whenever there is no need to distinguish a real from a complex manifold (just as in the analytic matter we intend to address). The given $V_n$  manifold can be considered as a hypersphere of an Euclidean space with $n+1$ dimensions that is represented in Cartesian coordinates by the equation
$$
\sum\limits_{\nu=1}^{n+1}  y^2_{\nu} = 1\, .
\leqno(16)
$$
In this case it might be convenient not to use $V_n$'s intrinsic coordinates, and rather imagine both the points and the directions as defined by their elements that determine the ambient space: the points by means of their Cartesian coordinates $y_{\nu}$ that are constrained by (16); the $(\alpha)$ directions by means of their direction cosines $\alpha_{\nu}$, that are constrained not only by
$$
\sum\limits_{\nu=1}^{n+1}  \alpha^2_{\nu} = 1\, ,
\leqno(17)
$$
\noindent but also by the condition
$$
\sum\limits_{\nu=1}^{n+1}  \alpha_{\nu}\, y_{\nu} = 0
\leqno(18)
$$
\noindent of belonging to $V_n$ (i.e., to the tangent hyperplane).
\par Given a geodesic line $C$ within $V_n$, let's examine how the $\alpha_{\nu}(s)$ must change along $C$ so that their parallel directions (in $V_n$) shall be identified. Therefore, let's consider again equation $(I)$ at paragraph 2 and we shall see that constraints (1) will now only consist in equation (16). Lagrange's classic procedure certainly allows us to replace $(I)$ with the explicit equations
$$
\alpha'_{\nu} = \mu \, y_{\nu} \qquad\qquad\qquad (\nu=1 , 2,\dots, n+1),
\leqno(19)
$$
\noindent where $\mu$ denotes a multiplier that is a priori undetermined.
\par This has to form a system with (18) and thus defining both $\alpha$ as well as $\mu$.
\vfil
\eject
\noindent Equation (17) is in fact consistent with equations (18) and (19), it follows that:
$$
 {{d}\over {ds}}\sum\limits_{\nu=1}^{n+1} \alpha^2_{\nu} = 2\, \sum\limits_{\nu=1}^{n+1} \alpha_{\nu}\, \alpha'_{\nu} =  2\,\mu\, \sum\limits_{\nu=1}^{n+1} \alpha_{\nu}\, y_{\nu} = 0 \, .
$$
\noindent Therefore, considering that $\psi$ is the angle between $(\alpha)$ and $C$,
$$
\cos \psi = \sum\limits_{\nu=1}^{n+1} \alpha_{\nu}\, y'_{\nu} \, ,
$$
\noindent and that (paragraph 8) such angle remains constant (as $s$ changes) based on the hypothesis that $C$ shall be geodetic.
\par Differentiating and considering equations (19) and (16), equation (18) yields:
$$
\mu + \cos \psi = 0 \, ,
$$
\noindent and therefore equation (19) takes the following form
$$
 \alpha'_{\nu} = - \cos \psi \,  y_{\nu} \, ,
$$
\noindent and the calculation of the $ \alpha_{\nu}$ cosines is reduced to quadratures. Actually this is not even necessary. First of all noticing that when $\cos \psi = 0$ (i.e., {\it for directions that are orthogonal to\/} $C$) then  $\alpha'_{\nu} =0$ will be sufficient. This means that {\it the characteristic of the parallelism in\/} $V_n$ {\it is that\/} $\alpha_{\nu}$ {\it are constant and therefore it coincides with the ordinary parallelism in the ambient space. For what concerns the directions that are not orthogonal to\/} $C$, the final terms of paragraph 6 can be considered in order to deduce that {\it the condition of parallelism (within $V_n$) consists in forming the same angles both with $C$ as well as with the  $n-1$ fixed (independent) directions that are orthogonal to $C$ and are of course tangent to the hypersphere\/} (16).
\par Specifically, let's consider $V_3$. The behaviour that has just been shown for the parallel directions coming orthogonally from the points of the same geodesic $C$ proves that there is no relation with the Clifford parallelism.
In fact, if $\infty^1$  Clifford parallel geodesics are drawn from the points of a $C$ geodesic, these will instead be normal to $C$ but their tangents will {\it not\/} be parallel within the ambient space
\footnote{($^{13}$)}{\foot{This can be justified with the following considerations. 
\par First of all let's consider  [L. Bianchi, loc. cit. ($^{1}$), p. 446] that the geodesics that form a Clifford congruence are parametrically described by formulae such as  
$$
y_{\nu} = a_{\nu}\, \cos s + \sin s \sum\limits_{\rho=1}^4 c_{\nu\rho} \, a_{\rho}\qquad\qquad\qquad (\nu=1 ,2,3,4), 
\leqno(1)
$$
\noindent where the parameter $s$ coincides with the arc and the constant coefficients $c_{\nu\rho}$ that identify the congruence have to satisfy both the ($c_{\nu\rho} + c_{\rho\nu}=0$) skew-symmetry condition as well as the orthogonality condition; the starting values for  $y_{\nu}$, i.e., $a_{\nu}$ are constrained by formula (16) and after all they vary (in all possible ways) from one of the many congruence geodesics to another. }}.
\smallskip
\centerline{11. {\bf Reduction of parallelism equations to the skew-symmetric type}}
\smallskip
The $(I_a)$ differential system allows (paragraph 6) a quadratic integral. Therefore
%
\vfil
\eject
\noindent it can be sustained  \footnote{}{\foot{\indent Let's multiply equations (1) by $c_{\mu\nu}$ and sum over index $\nu$ from $1$ to $4$. As expected, we get the following identities
$$
\sum\limits_{\nu=1}^{4}c_{\mu\nu} c_{\nu\rho}=-\sum\limits_{\nu=1}^{4}c_{\mu\nu} c_{\rho\nu}=-\varepsilon_{\mu\rho}\quad (\mu, \rho=1, 2, 3, 4;\,\varepsilon_{\mu\rho}=0\, {\rm for} \mu\not=\rho\, {\rm and}\, =1\,{\rm for}\,\mu=\rho),
$$
\noindent and then we immediately get  (as $\mu$ is changed into $\nu$ and $\nu$ into $\rho$ once the calculation has been carried out)
$$
\sum\limits_{\rho=1}^{4}c_{\nu\rho}y_{\rho}=\cos s \, \sum\limits_{\rho=1}^{4}c_{\nu\rho}a_{\rho} - a_{\nu}\, \sin s \qquad\qquad\qquad (\nu=1, 2, 3, 4).
\leqno(2)
$$
\indent The direction cosines of any curve (1) (in a generic point) are clearly represented by the derivatives of  $y_{\nu}$ with respect to the arc length $s$:
$$
y'_{\nu} = - a_{\nu}\, \sin s + \cos s \sum\limits_{\rho=1}^4 c_{\nu\rho} \, a_{\rho}\, ,
$$
\noindent which, thanks to equations (2), can be given as functions of the $y_{\nu}$ coordinates (of the point from which the congruence geodesic is drawn) in the following form
$$
y'_{\nu} = \sum\limits_{\rho=1}^4 c_{\nu\rho} \, y_{\rho} \qquad\qquad\qquad (\nu=1, 2, 3, 4).
$$
\indent As the $c_{\nu\rho}$ determinant does not vanish (its absolute value is $1$) this means that the cosines can {\it never\/} be the same in two different points. Therefore this excludes that there can be a curve $C$ from whose points Clifford congruence geodesics having the same direction can be drawn.\qquad {Q.E.D.}
}}\footnote{($^{14}$)}{\foot{Cf., e.g., Darboux, loc. cit.  ($^{6}$).}}  that the system itself can be reduced (by means of a suitable linear transformation of the unknowns) to the following skew-symmetric form
$$
{{d z_h}\over {ds}} = \sum\limits_{k=1}^{n} p_{hk}\, z_{k} \qquad\qquad\qquad (h=1, 2, \dots , n), 
\leqno(II)
$$
\noindent that is characterized by the following equations
$$
p_{hk} +p_{kh}=0\qquad\qquad\qquad\qquad (h, k=1, 2, \dots , n)
\leqno(20)
$$
\noindent between the coefficients (that anyhow are supposed to be any function of $s$). To this end it is sufficient that the linear transformation between $\xi^{(i)}$  and $z_h$ shall provide a canonical form to the quadratic integral
$$
\sum\limits_{h=1}^{n}z^2_h = {\rm const.}
$$
\noindent In order to achieve this and at the same time show the transformation from a geometrical point of view, it's best to proceed as follows.
\par Let's arbitrarily draw $n-1$ directions from each point $P$ of the curve $C$ that, with the one belonging to $C$, form an $\Omega$ orthogonal tuple. Let's assign the numbers $1, 2, \dots , n-1$ to these directions and identify the parameters (i.e., contravariant tensor tuple) with $\lambda^{(i)}_h$ ($i=1, 2, \dots , n$) (for the $h^{\rm th}$) and the moments with $\lambda_{h/i}$ (the covariant tensor tuple).

\vfil
\eject
\noindent In order to have a coherent notation, we will assign index $n$ to the direction of $C$ that completes the tuple. So we will set
$$
x'_i=\lambda^{(i)}_n\, ,\qquad \qquad \sum\limits_{j=1}^{n} a_{ij}\, x'_j=\lambda_{n/i}  \qquad\qquad (i=1, 2, \dots , n).
\leqno(21)
$$
\noindent Therefore the characteristic relations of the orthogonal tuples stand
$$
\sum\limits_{i=1}^{n} \lambda_{h/i}\, \lambda^{(i)}_k=\varepsilon_{hk}\qquad\qquad\qquad\qquad (h, k=1, 2, \dots , n)
\leqno(22)
$$
\noindent where (the symbol) $\varepsilon_{hk}$ has the usual meaning ($0$ when $h\not=k$ , $1$ when $h=k$).
\smallskip
\par Of course the $\Omega$ tuple (i.e., the tuple of orthonormal vectors that therefore are linearly independent) 
and the $\lambda$ directions through it have to be considered as known functions of $s$ (just as the $x_i$ 
and $x'_j$ functions).
\par Given the above, the linear transformation we are seeking is made explicit as follows. The $z_h$ cosines of the angles that are formed with the tuple's directions in each point are taken as the elements that determine the parallel directions in place of the $\xi^{(i)}$ parameters (vectors) (or of their corresponding $\xi_i$ elements (covectors)). The following geometric identity
$$
\sum\limits_{h=1}^{n}z^2_h = 1
$$
\noindent certainly ensures that a transformed skew-symmetric system shall be reached. Let's also carry out the calculation in order to have the explicit expression for the $p_{hk}$ coefficients.
\par By definition the following equations constrain the new $z_h$ unknowns to the $\xi^{(i)}$ parameters, or respectively to the $\xi_i$  parameters,
$$
z_h = \sum\limits_{i=1}^{n} \xi^{(i)}\lambda_{h/i}\, , \qquad
z_h = \sum\limits_{i=1}^{n} \xi_i \, \lambda^{(i)}_h  \qquad\qquad\qquad (h=1, 2, \dots , n).
\leqno(23_{a,b})
$$
\noindent Considering equations (22), they can be solved in the following way
$$
\xi^{(i)} = \sum\limits_{k=1}^{n} z_k  \, \lambda^{(i)}_k\, , \qquad
\xi_i = \sum\limits_{k=1}^{n} z_k \, \lambda_{k/i}  \qquad\qquad\qquad (i=1, 2, \dots , n).
\leqno(24_{a,b})
$$
Let's differentiate equations $(23_a)$ with respect to $s$ and consider equations $(I_a)$ [where, in accord with (21), $x'_j$ will be replaced by $\lambda^{(j)}_n$] then let's replace parameters $\xi^{(i)}$ (or $\xi^{(l)}$) with the $(24_a)$ values in the equation's left hand side (once the differentiation has been carried out). Where for the sake of brevity we shall put
$$
p_{hk} =  \sum\limits_{i=1}^{n}{ {d\lambda_{h/i}} \over {ds}} \,  \lambda^{(i)}_k \, -\,  \sum\limits_{i,j,l=1}^n\Bigl\{ ^{jl}_{\, i} \Bigr\} \,  \lambda^{(j)}_n\, \lambda^{(l)}_k\, \lambda_{h/i} \qquad\qquad (h, k=1, 2, \dots , n),
\leqno(25_{a})
$$
\noindent and we will have equations $(II)$. Nevertheless the values $(25_a)$ of the coefficients do not directly prove the skew-symmetric nature that, according to the previous observations, they must have.
\vfil
\eject
\noindent Transformations can be used to prove this when formulae (22) and the structure of the Christoffel symbols are considered. Nevertheless this can be achieved more easily by noting that these $z_h$ variable differential equations (i.e., equations $(II)$) can be obtained by applying the same method to the equivalent equations  $(23_b)$,  $(24_b)$,  $(I_b)$ in place of equations $(23_a)$,  $(24_a)$,  $(I_a)$. Carrying out the calculation in this way provides the following new expressions for the $p_{hk}$ coefficients
$$
p_{hk} =  \sum\limits_{i=1}^{n}{ {d \lambda^{(i)}_h} \over {ds}} \,  \lambda_{k/i} \, +\,  \sum\limits_{i,j,l=1}^n\Bigl\{ ^{ij}_{\, l} \Bigr\} \,  \lambda^{(j)}_n\, \lambda_{k/l} \, \lambda^{(i)}_h \qquad\qquad (h, k=1, 2, \dots , n).
\leqno(25_{b})
$$
\noindent The expected formal result is obtained by comparing expressions $(25_a)$ and $(25_b)$.

\par In fact, adding the expression for $p_{hk}$ that is provided by equations $(25_a)$ to the one for $p_{kh}$ given by equations $(25_b)$, the two $\sum\limits_{i,j,l=1}$ terms (addends) cancel out identically (this is clear when in one of them the indices $i$ and $l$ are interchanged) thus we are left with
$$
p_{hk} \, + \, p_{kh}=  \sum\limits_{i=1}^{n}\biggl( { {d\lambda_{h/i}} \over {ds}} \,  \lambda^{(i)}_k \, + \, { {d \lambda^{(i)}_k} \over {ds}} \,  \lambda_{h/i} \biggr)\, ,
$$
\noindent which also vanishes thanks to equations (22).
\smallskip
Linear system $(II)$ with skew determinant is an obvious generalization that can be seen in rigid body kinematics when the motion of the body-fixed axes has to be found given the angular velocity. With regard to the theory for such differential system, please see the works mentioned in the foreword [from note ($^{4}$) to ($^{7}$)]. Nevertheless we would like to draw the attention to some elementary properties:
\smallskip
\par $1^{\rm st}$ When an arbitrary orthogonal replacement (with coefficients that depend on $s$) is done for the $z_h$ unknowns, the transformed system is of the same kind . The demonstration is immediate if we realize that the quadratic integral $\sum\limits_{h=1}^{n}z^2_h = {\rm const.}$ [its existence is characteristic of the antisymmetric systems  $(II)$] preserves its form when the  $z_h$ unknowns undergo orthogonal transformations.

\smallskip
\par $2^{\rm nd}$ If the curve $C$ is geodesic, the $(I_a)$ parallelism equations are satisfied (cf. paragraph 7) by  $\xi^{(i)}=x'_i=\lambda^{(i)}_n$, therefore as a consequence [considering equations $(22)$ and $(23_a)$] equations $(II)$ are satisfied by
$$
z_h = 0 \qquad\qquad (h=1, 2, \dots , n-1), \qquad\qquad z_n = 1.
\leqno(26)
$$
\noindent This requires that
$$
\qquad p_{hn} = 0 \qquad\qquad\qquad (h=1, 2, \dots , n-1), 
\leqno(27)
$$
\noindent for any value of $s$. On the other hand, if (27) is true, then equation $(II)$ implies the solution (26) and the curve $C$ is geodesic. 
\smallskip
\par $3^{\rm rd}$ Always considering $C$ being a geodesic, the system is reduced, let's say, automatically once the particular solution (26) is known. In fact, considering equations (27) , the first $n-1$ equations $(II)$  do not contain any $z_n$ unknowns and the $n$-th equation is reduced to  ${ {d z_n} \over {ds}}=0 $.

\vfil
\eject
\noindent $4^{\rm th}$  When the $m$ independent parallel directions are known, i.e., the $m$ solutions that are independent from the initial $(I_a)$ equations, for arbitrary $C$, we know, from the linear systems' general theory,
that a reduction by $m$ units is possible. We want to prove that the actual reduction can also be achieved in another very simple way. And here is how.
\par Let's start by normalizing the $m$ known solutions (considering the final finding at paragraph 6) and derive the same number of mutually orthogonal ones (by means of rational algebraic operations). Then a new orthogonal tuple $\Omega^{\ast}$ (that includes the $m$ directions obtained from the known solutions' orthogonalization) will replace the initial tuple of the $\Omega$ directions that is linked to each point belonging to $C$ (that includes the direction of $C$ and the other $n-1$ ones that are orthogonal to it).
\par The $z_h^{\ast}$  direction cosines (of any direction) with respect to $\Omega^{\ast}$ are thus linked to the $z_h$ direction cosines (that refer to $\Omega$) by an orthogonal transformation. Once system $(II)$  has been transformed in to the $z_h^{\ast}$  variables (that remains antisymmetric), it will allow (by construction) $m$ distinct solutions of the $n-1$ kind for the null $z^{\ast}$  and the $n^{\rm th}$ will be equal to the unity. Assuming that an orthogonal transformation of this kind shall hold, for these $m$ $z_n^{\ast}, z_{n-1}^{\ast}, \dots , z_{n-m+1}^{\ast}$ solutions every $p_{hk}$ will be null when
$$
h=1, 2, \dots, n-m, \qquad k=n-m+1, n-m+2, \dots, n \, ,
$$
\noindent Therefore the first $n-m$ transformed equations form a reduced system (that are also skew-symmetric) that includes the $z_1^{\ast}, z_2^{\ast}, \dots , z_{n-m}^{\ast}$  unknowns. \qquad {Q.E.D.}
\bigskip
\centerline{12. {\bf Three-dimensional spaces}}
\medskip
When $n=3$, the system $(II)$ is exactly that which the moving trihedron theory relies on \footnote{($^{15}$)}{
\foot{G.~Darboux,~}\foots{Le{\c c}ons sur la th{\` e}orie g{\` e}nerale des surfaces}\foot{,~$2^{\rm e}$ {\' e}dition, t. I (Paris, Gauthier-Villars, 1914), Chap. II, pp. 27-41.}}: all that has to be done is to provide a different interpretation of the results (in regard to the parallelism in a $V_3$ along a given curve $C$). Let's keep to the most simple case where $C$ is geodesic. Following point 3 in the previous paragraph, in this case the $z_1, z_2, z_3$ variable differential system $(II)$ is reduced (when $z_3 = {\rm const.}$, associated) to the following two-equation system
$$
{{d z_1 }\over {ds}}=p_{12}\, z_2\, , \qquad {{d z_2 }\over {ds}}=p_{21}\, z_1\, .
$$
\noindent From the equation $p_{12} + p_{21}=0$ we then have integral $z_1^2 + z_2^2={\rm const.}$ and once the (integration) constant is named as $r^2$ we can set $r=\cos \vartheta\,$, and $r=\sin \vartheta\,$. Therefore we have a single equation for $\vartheta$
$$
{{d \vartheta}\over {ds}}=p \qquad\qquad\qquad (p=-p_{12}=p_{21})\, .
$$
\vfil
\eject
\noindent This leads to a simple quadrature. 
\bigskip
\centerline{13. {\bf Relation between the $p_{hk}$ coefficients and the Ricci rotation coefficients}}
\medskip
Let's suppose that curve $C$ shall belong to a congruence of curves (assigned but unspecified) in $V_n$. Based on this hypothesis, the $\lambda^{(i)}_n$ and the $\lambda_{n/i} $ parameters can be considered as functions of the $x_1, x_2, \dots , x_n$ coordinates that (along $C$ and thanks to equations (3)) are reduced to the functions of $s$ that have been considered earlier (paragraph 11).
\par Let's suppose we associate $n-1$ other congruences to this one which the curve $C$ belongs to, so that we get an orthogonal tuple (these are the tangent vectors to it). The only condition that needs satisfying is that it shall coincide with $\Omega$ in the points belonging to $C$. Let the parameters $\lambda^{(i)}_h(x_1, x_2, \dots , x_n)$, $\lambda_{h/i}(x_1, x_2, \dots , x_n)$  be the corresponding contravariant and covariant coordinate systems.
\par It follows that, along $C$,
$$
{ {d\lambda_{h/i}} \over {ds}} = \sum\limits_{j=1}^{n}{ {\partial\lambda_{h/i}} \over {\partial x_j}}\, x'_j = \sum\limits_{j=1}^{n}{ {\partial\lambda_{h/i}} \over {\partial x_j}}\, \lambda^{(j)}_n \, ,
$$
\noindent so that equation $(25_a)$ (interchanging the two indices of summation $i$ and $l$ in the second term) can be written as:
$$
p_{hk} =  \sum\limits_{i,j=1}^{n} \lambda^{(i)}_k\, \lambda^{(j)}_n\, \biggl[	
{ {\partial\lambda_{h/i}} \over {\partial x_j}} \, -\,  \sum\limits_{l=1}^n\Bigl\{ ^{ij}_{\, l} \Bigr\}\, \lambda_{h/l} \biggr] \, .
$$ 
\noindent We can recognise the covariant derivative \footnote{($^{16}$)}{\foot{Cf. G.~Ricci et T.~Levi-Civita,~loc. cit. ($^{11}$).}} $\lambda_{h/ij}$ in the quantity within the parenthesis thus, if we
recall the formulae that define the following {\it rotation coefficients\/},
$$
\gamma_{hkn} =  \sum\limits_{i,j=1}^{n}\lambda_{h/ij}\, \lambda^{(i)}_k\, \lambda^{(j)}_n\, , 
$$ 
\noindent we can conclude that
$$
p_{hk} = \gamma_{hkn} \qquad\qquad (h, k=1, 2, \dots , n).
$$
\noindent Now we know that the following interpretation stands for $\gamma_{hkn}\, ds$ (being a rotation coefficient): when the $\Omega$ tuple's origin moves a distance $ds$ along $C$ from a generic point $P$ to a very close point $P'$, then the direction $h$ \footnote{($^{17}$)}{\foot{Which is the direction that is defined by the $\lambda^{(i)}_h$ parameters (or by the $\lambda_{h/i}$ moments).}} in $P'$ generally will not remain orthogonal to the $k$ direction that corresponds to point $P$ but it will form a ${\pi \over {2}} - \gamma_{hkn}\, ds$ angle with it.
\smallskip
\par The same identical interpretation is therefore possible for $p_{hk}\, ds$. From this
geometrical point of view,
\vfil
\eject
\noindent we can see in systems $(II)$ a generalization of the moving trihedron's elementary theory.
\bigskip
\centerline{14. {\bf Manifold with a congruence of parallel curves with respect to any traverse}}
\medskip
Let  $\xi_i (x_1, x_2, \dots , x_n)$ be the covariant coordinate system of congruence of curves in $V_n$.
Let's suppose that the curves belonging to this congruence are unconditionally parallel, i.e., they satisfy the conditions of parallelism along any line $C$. Specifically, let's choose a congruence line passing through $C$. Clearly, this congruence is formed by (curved) lines that must be geodesic (cf. paragraph 7). In general, this means that the parallelism equations (in whatever form they shall be given), e.g., equations $(I_b)$, should be verified every point belonging to $V_n$ \footnote{($^{18}$)}{\foot{As usual, to be understood as the field of $V_n$ that is being considered (where weak qualitative limits are understood).}}  and for any $x'_j$ direction.
\smallskip
\noindent Thus, making ${ {d\xi_{i}} \over {ds}}$ explicit, equations $(I_b)$ shall be written as:
$$
\sum\limits_{j=1}^{n} x'_j\, \biggl[	
{ {\partial\xi_{i}} \over {\partial x_j}} \, -\,  \sum\limits_{l=1}^n\Bigl\{ ^{ij}_{\, l} \Bigr\}\, \xi_{l} \biggr] =0 \qquad\qquad (i=1, 2, \dots , n),
$$ 
\noindent the $\xi_i$ components must satisfy the $n^2$ equations
$$
{ {\partial\xi_{i}} \over {\partial x_j}} \, -\,  \sum\limits_{l=1}^n\Bigl\{ ^{ij}_{\, l} \Bigr\}\, \xi_{l} =0 \qquad\qquad (i, j=1, 2, \dots , n),
\leqno(28)
$$ 
\noindent which shows that the $\xi_{ij}$ covariant (tensor) system ---that is the covariant derivative of the $\xi_i$ tensor--- vanishes identically.
\par
Specifically equations (28) lead to
$$
{ {\partial\xi_{i}} \over {\partial x_j}} = { {\partial\xi_{j}} \over {\partial x_i}}\, ,
$$
\noindent hence we see that the $\xi_i$ components are the partial derivatives of the same function $F$. Therefore equation 13 or, if you prefer, the equivalent formula
$$
\sum\limits_{i,j=1}^n a^{(ij)} \, \xi_i  \, \xi_j  = 1\, ,
$$
\noindent becomes
$$
\Delta F = 1 \qquad \hbox{{\rm (where $\Delta$ is a differential parameter of first order)}},
$$
\noindent and proves that the $F={\rm const.}$ hypersurfaces (where the curves belonging to our congruence are its orthogonal trajectories) are geodesically parallel.
\vfil
\eject
As a matter of simplicity, let the function $F$ be the $x_n$ coordinate and let's associate with it (as $x_1, x_2, \dots , x_{n-1}$ coordinates) the $n-1$ independent integrals of the following equation
$$
\nabla (x_n, \Theta) = 0 \qquad \hbox{{\rm (where $\nabla$ is a mixed differential parameter)}},
$$
\noindent thus the $x_h = {\rm const.}$ $(h=1, 2, \dots , n-1)$ coordinate hypersurfaces are orthogonal to the $x_n = {\rm const.}$ coordinates. In turn, we can write
$$
a^{(nj)}=0 \qquad (j=1, 2, \dots , n-1), \qquad a^{(nn)}=1 \, ,
\leqno(29)
$$
\noindent and 
$$
\xi_j=0 \qquad (j=1, 2, \dots , n-1), \qquad \xi_n=1 \, ;
$$
\noindent and then equations (28) will be reduced to
$$
\eqalignno{
\Bigl\{ ^{ij}_{\, n} \Bigr\}=0 & \qquad & (i, j=1, 2, \dots , n),
}
$$ 
\noindent and even, considering the following identities
$$
\Bigl\{ ^{ij}_{\, n} \Bigr\} = \sum\limits_{h=1}^n  a^{(hn)} \, a_{ij,h}
$$
\noindent as well as equations (29), it follows that:
$$
\eqalignno{
a_{ij,n}=0 & \qquad & (i, j=1, 2, \dots , n).
}
$$
\noindent We note that equations (29) are equivalent to the analogous ones relating to their reciprocal
elements:
$$
a_{nj}=0 \qquad (j=1, 2, \dots , n-1), \qquad a_{(nn)}=1 \, ,
\leqno(29')
$$
\noindent and bearing in mind that
$$
a_{ij,n} = {1 \over 2}\biggl( {{\partial a_{in}} \over {\partial x_{j}}} + {{\partial a_{nj}} \over {\partial x_{i}}}  - {{\partial a_{ij}} \over {\partial x_{n}}}  \biggr)\, ,
$$
\noindent we finally get
$$
{ {\partial a_{ij}} \over {\partial x_n}} =  0  \quad \qquad  (i, j=1, 2, \dots , n-1). 
\leqno(28')
$$
\noindent Equations $(29')$ and $(28')$ show us that $d s^2$ {\it takes the form\/}
$$
d x_n^2 \, + \, d \sigma^2 \, ,
\leqno(30)
$$
\noindent where $d \sigma^2$ is the square of an arbitrary line element with $n-1\,$ variables $x_1, x_2, \dots , x_{n-1}$ (whose coefficients are independent of $x_n$).
\par
Given that \footnote{($^{19}$)}{\foot{Cf.~J.~Hadamard,~}\foots{Sur les {\' e}l{\' e}ments lin{\' e}aires \`a plusieurs dimensions,~}\foot{Bulletin des Sciences Math{\' e}ma\-tiques, t. XXV (1901), pp. 37-40.~}} the form of equation (30) is characteristic for the manifolds that allow a simple infinity of {\it geodesic surfaces\/} \footnote{($^{20}$)}{\foot{Any $n-1$ dimensional manifolds that are immersed in a $V_n$ manifold are called geodesic surfaces when they contain the entire geodesic of $V_n$ that joins any two points belonging them.}}  , then we conclude that
\vfil
\eject
\noindent {\it for a $V_n$ manifold the two properties (admitting ${\infty}^1$ geodesic surfaces and holding a complete parallelism congruence) shall always coincide.\/}
\bigskip
\centerline{15. {\bf Second order differentials –-- Invariant determinations –-- Ricci's lemma}}
\medskip
Let's consider the independent variables, e.g., $x_1, x_2, \dots , x_{n}$, to be fixed for a given investigation. As we have seen from the calculation, second order differentials $d^2 x_1, d^2 x_2, \dots ,d^2 x_{n}$ can always be considered null. However, the nature of such a convention is not invariant in regard to the changing of the variables. In fact, if we replace the $x_i$ variables with $n$ of their ${\bf x}_i( x_1, x_2, \dots , x_{n})$ independent combinations, usually the second differentials
$$
d^2 \, {\bf x}_i = \sum\limits_{j,l=1}^n  {{\partial^2  {\bf x}_{i}} \over {\partial x_j \partial x_l}} \, d x_j \, d x_l 
$$
\noindent (calculated supposing that $d^2 x_i=0$) shall not vanish.
\par
An invariant characterization can be easily reached when a quadratic differential form (provided that a metric of a $V_n$ is considered and the notations of the previous paragraphs are used) is assigned to the variables. It is sufficient to take the $d^2 x_i$ (not vanishing, but) defined as:
$$
d^2 x_i \, + \, \sum\limits_{j,l=1}^n\Bigl\{ ^{jl}_{\, i} \Bigr\} \, d x_j  \, d x_l   = 0 \qquad\qquad\qquad (i=1 ,2,\dots,n).
$$
\par
Once the geodesic equations (paragraph 7) are multiplied by $d s^2$, it will be clear that these $d^2 x_i$ are the ones that belong to the variables along the geodesic through a generic point $(x_1, x_2, \dots , x_{n})$ and along a $(d x_1,d  x_2, \dots , d x_{n})$ direction that is also generic. Such geometric interpretation assures {\it a priori\/} that the above convention has the required invariant nature (without any need of a material check that anyhow could easily be carried out).
\smallskip
\par
This is also true for the superposition of two independent systems that are incremented by $d x_i$ and $\delta x_i$. One could set $d \, \delta x_i = \delta \, d x_i =0$, but as can easily be seen, while the nature of the invertibility of increments $d$ and $\delta$ is invariant (as can easily be checked), the same does not hold when $d \, \delta x_i  =0$ is set. Let's replace them with:
$$
d \, \delta x_i \, + \, \sum\limits_{j,l=1}^n\Bigl\{ ^{jl}_{\, i} \Bigr\} \, d x_j  \, \delta x_l   = 0\, ,
\leqno(31)
$$
\noindent that imply
$$
d \, \delta x_i = \delta \, d x_i  \, ,
\leqno(31')
$$
\noindent and include (as a particular case when $d=\delta$) the previous expressions for the $d^2 x_i \,$.
\smallskip
\par
Regarding the change of variables, the invariance of equations (31) in this case can still be inferred from the geometrical interpretation. Observing that setting $\delta x_i = \varepsilon \, \xi^{(i)}$ (being $\varepsilon$ an infinitesimal constant) is sufficient so that equations (31) shall become identical to equations $(I_a)$. Thus they show how the displacements $(d x_1, d x_2, \dots , d x_{n})$ will cause the increments $\delta x_i$ to change,
\vfil
\eject
\noindent so that mutually parallel directions can be defined. As well as verifying this invariant property directly, it can also be checked by using an elegant formal technique suggested indirectly by Riemann \footnote{($^{21}$)}{\foot{loc.~cit. ($^2$),~}\foot{p. 381.~}} and made explicit by Weber \footnote{($^{22}$)}{\foot{loc.~cit. ($^2$),~}\foot{p. 388.~}}.
\smallskip
\par
From equations (31) and bearing in mind that
$$
d \, a_{ik} = \sum\limits_{j=1}^n  {{\partial {a_{ik}}} \over {\partial x_j}} \, d x_j = {1 \over 2} \,  \sum\limits_{j=1}^n \bigl( a_{ij,k} \, + \, a_{jk,i} \bigr) \, d x_j = {1 \over 2} \, 
\sum\limits_{j, l=1}^{n} \biggl[ a_{lk}\, \Bigl\{ ^{ij}_{\, l} \Bigr\}  \, +\,  a_{li} \, \Bigl\{ ^{jk}_{\, l} \Bigr\} \biggr]\, \, d x_j \, ,
$$
\noindent it follows identically
$$
d \sum\limits_{i,k=1}^n a_{ik} \, \delta x_{i} \, \delta x_{k} = 0 \, ,
\leqno(32)
$$
\noindent as well as
$$
d \sum\limits_{i,k=1}^n a_{ik} \, d x_{i} \, \delta x_{k} = 0 \, .
$$
\par
These equations correspond to the well-known result of the absolute differential calculus wherein the fundamental form's covariant derived system of the coefficients $a_{ik}$ vanishes identically (Ricci's lemma).
\bigskip
\centerline{16. {\bf Higher-order differentials. Invariant that summarises Riemann's symbols}}
\centerline{{\bf and provides the explicit expression in the most direct way}}
\medskip
If we apply equations (31) repeatedly, also the higher-order differentials of the $d' \, \delta \, d x_{i}\,$, $\delta \, d' \, d x_{i}$ kind (using of course symbol $d'$ the same way as $d$ and $\delta$) will surely be defined (as functions of the first order differentials of the Christoffel symbols and their derivatives).
\par
We can't state that $d' \, \delta \, d x_{i}$ shall coincide with $\delta \, d' \, d x_{i}$. On the contrary, the actual calculation
indeed proves that they differ. However, the following differences
$$
u^{(i)} = d' \, \delta \, d x_{i} \, - \, \delta \, d' \, d x_{i}
\leqno(33)
$$
\noindent form a contravariant system (i.e., a vector) which is surely a remarkable feature.
\par
In order to prove this, let's consider an auxiliary covariant system $p_i$ consisting in $n$ functions of the $x$ variables (without making use of any differentials) and first of all let's note (as we suppose that system $p_i$ shall be covariant) that
$$
\sum\limits_{i=1}^n p_i \, d x_{i} 
$$
\vfil
\eject
\noindent is an invariant. As both operators $d'$ and $\delta$ are invariants this is also true for their difference:
$$
G = d' \, \delta \sum\limits_{i=1}^n p_i \, d x_{i} \, - \, \delta \, d' \sum\limits_{i=1}^n p_i \, d x_{i}  = 
d' \sum\limits_{i=1}^n \bigl( \delta \, p_i \, d x_{i}\, + \, p_i \, \delta \, d x_{i}\bigr) \, - \, \delta \sum\limits_{i=1}^n \bigl( d' \, p_i \, d x_{i}\, + \, p_i \, d' \, d x_{i}\bigr),
$$
\noindent that we will expand bearing in mind that (thanks to equations $(31')$) we have
$$
d' \, \delta p_i = \delta \, d' p_i  \, ,
$$
\noindent thus we get
$$
G = \sum\limits_{i=1}^n p_i \bigl( d' \, \delta \, d x_{i} \, - \, \delta \, d' \, d x_{i} \bigr) =  \sum\limits_{i=1}^n p_i \, u^{(i)}\, ,
$$
\noindent hence we can see that also the term $ \sum\limits_{i=1}^n p_i \, u^{(i)}$ is an invariant. From this invariance and that the $p_i$ coefficients (functions of $x$ having the only constraint of transforming with a covariant law when a variable change is carried out) are arbitrary leads to the $u^{(i)}$ system (defined by equations (33)) being contravariant. \qquad {Q.E.D.}
\smallskip
\par
This shows that equation:
$$
I = \sum\limits_{i,k=1}^n a_{ik} \,  u^{(i)} \,  \delta' x_{k} \, ,
\leqno(34)
$$
\noindent is invariant, where $\delta' x_{k}$ is an arbitrary system of increments of the variables.
\par
From equations (31)
$$
\displaylines{ d' \, \delta \, d x_i  = - d' \, \sum\limits_{j,l=1}^n\Bigl\{ ^{jl}_{\, i} \Bigr\} \, d x_j  \, \delta x_l  
 \cr = - \,  \, \sum\limits_{j,l,h=1}^n { {\partial \Bigl\{ ^{jl}_{\, i} \Bigr\}} \over {\partial x_h}} \, d' x_h\, d x_j  \, \delta x_l  - \, \sum\limits_{j,l=1}^n\Bigl\{ ^{jl}_{\, i} \Bigr\} \bigl( d' \, dx_j \, \delta x_l \, + \, dx_j \, d' \, \delta x_l \bigr),
 \cr \delta \, d' \, d x_i  = - \delta \, \sum\limits_{j,l=1}^n\Bigl\{ ^{jl}_{\, i} \Bigr\} \, d x_j  \, d'  x_l  
\cr = - \,  \, \sum\limits_{j,l,h=1}^n { {\partial \Bigl\{ ^{jl}_{\, i} \Bigr\}} \over {\partial x_h}} \, \delta x_h\, d x_j  \, d' x_l  - \, \sum\limits_{j,l=1}^n\Bigl\{ ^{jl}_{\, i} \Bigr\} \bigl( \delta \, dx_j \, d' x_l \, + \, dx_j \, \delta \, d' x_l \bigr),
}
$$
\noindent and by subtraction (interchanging indices $h$ and $l$ in the first summation) we get
$$ 
 \displaylines{ u^{(i)} = d' \, \delta \, d x_{i} \, - \, \delta \, d' \, d x_{i} 
\cr  = \sum\limits_{j,l,h=1}^n 
\Biggl[{ {\partial \Bigl\{ ^{jl}_{\, i} \Bigr\}} \over {\partial x_h}}  \, - \,
{ {\partial \Bigl\{ ^{jh}_{\, i} \Bigr\}} \over {\partial x_l}}\Biggr]
\, dx_j\, d' x_l  \, \delta x_h  - \, \sum\limits_{j,l=1}^n\Bigl\{ ^{jl}_{\, i} \Bigr\} \bigl( d'  \, dx_j \, \delta x_l \, - \, 
\delta \, dx_j \, d' x_l \bigr).
}
$$
\noindent From (31) we have
$$
d' \,  d x_j  = - \, \sum\limits_{j,l=1}^n\Bigl\{ ^{ht}_{\, j} \Bigr\} \, d x_h  \, d'  x_t \, , \qquad 
\delta \,  d x_j  = - \, \sum\limits_{j,l=1}^n\Bigl\{ ^{ht}_{\, j} \Bigr\} \, d x_h  \, \delta  x_t \, , 
$$
\noindent so that, after an exchange of indices, we can take out the common factor $d x_j  \, d' x_l \, \delta  x_h$,
\vfil
\eject
\noindent we see that 
$$
 - \, \sum\limits_{j,l=1}^n\Bigl\{ ^{jl}_{\, i} \Bigr\} \bigl( d'  \, dx_j \, \delta x_l \, - \, 
\delta \, dx_j \, d' x_l \bigr)
= \sum\limits_{j,l,h=1}^n dx_j\, d' x_l  \, \delta x_h \,
\sum\limits_{t=1}^n \Biggl[
\Bigl\{ ^{th}_{\, i} \Bigr\} \, \Bigl\{ ^{jl}_{\, t} \Bigr\}
\, - \, \Bigl\{ ^{tl}_{\, i} \Bigr\} \, \Bigl\{ ^{jh}_{\, t} \Bigr\}
\Biggr]\, ,
$$
\noindent and we finally get
$$
u^{(i)} = \sum\limits_{j,l,h=1}^n dx_j\, d' x_l  \, \delta x_h  \bigl\{j i,\, l h \bigr\} \, ,
\leqno(35)
$$
\noindent where we denote
$$
\bigl\{j i,\, l h \bigr\} =
{ {\partial \Bigl\{ ^{jl}_{\, i} \Bigr\}} \over {\partial x_h}}  \, - \,
{ {\partial \Bigl\{ ^{jh}_{\, i} \Bigr\}} \over {\partial x_l}}
\, + \,
\sum\limits_{t=1}^n \Biggl[
\Bigl\{ ^{jl}_{\, t} \Bigr\} \, \Bigl\{ ^{th}_{\, i} \Bigr\}
\, - \, \Bigl\{ ^{jh}_{\, t} \Bigr\} \, \Bigl\{ ^{tl}_{\, i} \Bigr\}
\Biggr]
\leqno(36)
$$
\noindent Riemann's symbols of the second kind.
\smallskip
\par
If we now turn to those of the first kind, by setting
$$
a_{jk,lh} = \sum\limits_{i=1}^ n a_{ik} \bigl\{j i,\, l h \bigr\} \, ,
\leqno(37)
$$
\noindent considering equations (35), expression (34) of $I$ becomes
$$
I = \sum\limits_{j,l,h,k=1}^n a_{jk,lh} \, dx_j\, d' x_l  \, \delta x_h \, \delta' x_k
\leqno(34')
$$
\noindent and directly highlights the covariant nature of symbols (37) and, as far as I can see, this is the
fastest way to reach this result. These symbols' properties can be summarised in
the following formulae:
$$ 
 \displaylines{ a_{jk,lh} = - \, a_{jk,hl} \, ,
\cr  a_{jk,lh} = a_{lh,jk} \, ,
}
$$
\noindent and all we need to do is to refer to their ordinary use and consider the definition formulae to deduce them (the first one immediately and the second one by means of some transformations).
\par
Instead, for the rather important case where the independent differentials are reduced to two (when $d'$ coincides with $d$ and $\delta'$ with $\delta$) we shall have to focus on the geometrical interpretation of invariant $I$.
\bigskip
\centerline{17. {\bf Parallelogramoids --- Base and suprabase.}}
\centerline{{\bf Vertices coordinates development starting from the base}}
\medskip
Let $PQ$ be a generic geodetic arc belonging to manifold $V_n$. Let's suppose we draw two more geodesics along parallel directions from points $P$ and $Q$. They will form (cf. paragraph 8) the same angle $\psi$ with $PQ$. Let's take two identical arcs on these geodesics
$$
PP' = QQ' = d s \, ,
$$
\noindent and let's also join $P'$ and $Q'$ with a geodesic arc. Thus we will have a geodesic
\vfil
\eject
\noindent quadrangle that we shall call {\it parallelogramoid\/} and we will name {\it base\/} and {\it suprabase\/} the two opposite sides $PQ$ and $P'Q'$.
\par
Let's identify the coordinates of the four vertices as $P$, $Q$, $P'$, $Q'$ with $x_i^{(P)}$, $x_i^{(Q)}$, $x_i^{(P')}$,$x_i^{(Q')}$, the direction parameters of the two parallel geodesics in their origins with $\xi_P^{(i)}$, $\xi_Q^{(i)}$ and the values of the Christoffel symbols in these points with $\Bigl\{ ^{jl}_{\, i} \Bigr\}_P$, $\Bigl\{ ^{jl}_{\, i} \Bigr\}_Q$.
\par
Excluding the terms in $d s$ that are higher than the second order, the geodesic equations
provide
$$
\cases{
x_i^{(P')} = x_i^{(P)} \, + \, d s\, \xi_P^{(i)} \, - \, {1 \over 2} \, d s^2 \sum\limits_{j,l=1}^n
\Bigl\{ ^{jl}_{\, i} \Bigr\}_P \, \xi_P^{(j)} \, \xi_P^{(l)} \, , \cr
x_i^{(Q')} = x_i^{(Q)} \, + \, d s\, \xi_Q^{(i)} \, - \, {1 \over 2} \, d s^2 \sum\limits_{j,l=1}^n
\Bigl\{ ^{jl}_{\, i} \Bigr\}_Q \, \xi_Q^{(j)} \, \xi_Q^{(l)} \qquad \qquad (i = 1, 2, \dots , n).\cr
}
\leqno(38)
$$
\noindent Denote by $\delta x_i$ the difference $x_i^{(Q)} - x_i^{(P)}$ between points $P$ and $Q$ (and in general by $\delta f$ the difference of any element $f$, whether it's numeric or geometric) . And subtracting equations (38) we will get
$$
x_i^{(Q')} - x_i^{(P')} = \delta x_i \, + \, d s \, \delta \, \xi^{(i)} \, - \, {1 \over 2} \, d s^2 \, \delta \sum\limits_{j,l=1}^n
\Bigl\{ ^{jl}_{\, i} \Bigr\} \, \xi^{(j)} \, \xi^{(l)} \, .
$$
\noindent Considering that by construction $d s$ shall not change from $P$ to $Q$ and considering it as infinitesimal, setting also for the sake of brevity
$$
d x_i = \xi^{(i)} d s \, ,
$$
\noindent and taking into account equation (31) with $\delta = d$, we can also write
$$
D x_i = x_i^{(Q')} - x_i^{(P')} =
\delta x_i \, + \, \delta \, dx_i\, + \, {1 \over 2} \, \delta\,  d^2 x_i   \qquad \qquad (i = 1, 2, \dots , n).
\leqno(39)
$$
\par
From the above we see that the $D x_i$ differences of the homologous coordinates of points $P'$ and $Q'$ are (considering the accepted approximation regarding $d s$) of the first order in regard to the $\delta$ variation. This means that if we identify the $PQ$ arc with $\delta s$ and we treat it as infinitesimal (independent from $d s$) then the $D x_i$ differences will also be infinitesimals of the first order (at least) in regard to $\delta s$. Therefore equations (39) provide an expression for it that is exact:
\smallskip
\par
{\it up to the second order for\/} $d s$;
\par
{\it up to the first order for\/} $\delta s$.
\smallskip
\par From the meaning of symbols $d$ and $\delta$ (as they appear in equation (39)) we immediately see
(as also for $d^2$) that also $\delta d$ and $\delta d^2$ are made explicit thanks to equation (31). And it's hardly necessary to warn that once the calculations have been carried out, all the functions of the place must refer to point $P$.
\bigskip
\centerline{18. {\bf Suprabase length –-- Curvature}}
\medskip
Let's name the coefficients of the square of the line element in $P'$ as $a'_{ik}$.
\vfil
\eject
\noindent In the light of what we have just seen about the differences $D x_i$, we can consider
$$
\sum\limits_{i,k=1}^n a'_{ik}\, D x_i \, D x_k
\leqno(40)
$$
\noindent as an {\it expression of the square of the $\overline{P'Q'}^2$ distance, as long as the terms of order higher than the second, both in $d s$ as well as in $\delta s$, are neglected\/}.
\par
Calculating also the values for $a'_{ik}$ in point $P$ (with the same approximation) will be helpful. To this end it's sufficient to develop (neglecting the $ds^3$ terms) along the $PP'$ geodesic; thus we have
$$
a'_{ik} = a_{ik} + d a_{ik} + \, {1 \over 2} \, d^2 a_{ik} \, .
\leqno(41)
$$
A similar transport can be found in a generic $D x_i $ once the ${1 \over 2} \, d^2 \delta x_i $ term is added and subtracted (and bearing in mind that $\delta \, d x_i  = d \, \delta x_i$). Equations (39) become
$$
D x_i = D' \delta x_i \, - \, {1 \over 2} \, v^{(i)} \, ,  
\leqno(39')
$$
\noindent where
$$
\leqalignno{
D' \delta x_i & = 
\delta x_i \, + \, d \, \delta x_i\, + \, {1 \over 2} \, d^2 \delta x_i  \, ,  
& (42) \cr
v^{(i)}  &= d \, \delta \, dx_i\, - \, \delta \, d^2 x_i    \qquad \qquad (i = 1, 2, \dots , n). 
& (43) \cr
}
$$
\noindent Clearly equations $v^{(i)}$ are of the third order (the second is in $d s$ and the first in $\delta s$). Therefore replacing in equation (40) (neglecting the terms of an overall order that is higher than the fourth), we have
$$
\overline{P'Q'}^2 = \sum\limits_{i,k=1}^n a'_{ik}\, D' \delta x_i \, D' \delta x_k \, - \,
\sum\limits_{i,k=1}^n a'_{ik}\, v^{(i)} D' \delta x_k \, .
$$
\noindent Using equations (41) and (42) (and with the same approximation), the first summation can be put in the following form
$$
\delta s^2\, + \, d \, \delta s^2\, + \, {1 \over 2} \, d^2 \delta s^2  \, ,  
$$
\noindent where
$$
\delta s^2 = \overline{PQ}^2 = \sum\limits_{i,k=1}^n a_{ik}\, \delta x_i \, \delta x_k \, ;
$$
\noindent and the second summation can be reduced to
$$
J = \sum\limits_{i,k=1}^n a_{ik}\, v^{(i)} \, \delta x_k \, .
$$
\par
Now, thanks to equation (32), $d \, \delta s^2$ vanishes identically; thus this is also true for $d^2 \delta s^2 $. Therefore
$$
\overline{P'Q'}^2 = \overline{PQ}^2 - J\, ,
\leqno(44)
$$
\noindent but we still have to provide a final expression for $J$. This is due to the circumstance that (in accord with equations (43)) $J$ can be considered a particular case of the $I$ invariant that is defined by
equation (34): setting $d'$ to coincide with $d$ and $\delta '$ with $\delta$ will be sufficient. Therefore we get
\vfil
\eject
\noindent from equation $(34')$
$$
J = \sum\limits_{j,l,h,k=1}^n a_{jk, lh}\, dx_j \, dx_l \, \delta x_h \, \delta x_k \, .
\leqno(45)
$$
\par
Now we can provide an intrinsic characterisation of the curvature according to the following
geometrical form:
\par
{\it Let's construct an infinitesimal parallelogramoid $PQP'Q'$ in the $V_n$ manifold and let's consider the ratio\/}
$$
K = {{ \overline{PQ}^2 - \overline{P'Q'}^2}\over {(ds\, \delta s \,\sin \psi)^2}} 
\leqno(46)
$$
\noindent {\it between the difference of the squares of the base and of the suprabase and the square of the parallelogramoid's area\/} \footnote{($^{23}$)}{\foot{More precisely, of any portion of a two-dimensional surface having the parallelogramoid as its contour and that approaches zero with it.}}. {\it This ratio (thanks to equation (44))\/}
$$
K = {{ J}\over {(ds\, \delta s \,\sin \psi)^2}} 
\leqno(47)
$$
\noindent {\it represents the Riemannian curvature of the manifold $V_n$ at the point $P$ on the parallelogramoid's plane.\/} The correspondence with the ordinary expression \footnote{($^{24}$)}{\foot{L.~Bianchi, loc.~cit. ($^1$),~}\foot{pp. 341-342.~}} follows from equation $(45)$.
\par
Once the ratio $K$ has been found to be depending only on point $P$ and on the plane, equation (46) provides the following corollary: all equivalent parallelogramoids that lie on the same base (and belong to the same plane) have equal length suprabases.
\par
I would also like to point out that (from a systematic point of view) the procedure that has just been carried out lays down a new property for the Riemannian curvature as well as offering the advantage of avoiding (when compared to the ordinary use) some formal developments. In fact, usually the curvature of the $V_2$ manifolds is defined first (following one of the ways put forward by Gauss); then we get the $V_n$ manifolds by using the $V_2$ manifolds that are obtained by the geodesics that belong to the same plane. Therefore quite some calculation is required in order to recognise that
equation (47) provides the Riemannian curvature for these $V_2$ manifolds.
\par
If instead, geometrical definition (46) is used for the $K$ ratio, then on the one hand the analytical translation that leads to equation (47) is more effective and on the other (obviously standing the same definition also for $n=2$) it will be clear that the curvature of manifold $V_n$ that is deduced from a generic parallelogramoid will coincide with that of any $V_2$ manifold (which, at the limit, can contain the parallelogramoid itself).
\bigskip
\centerline{CRITICAL NOTE}
\medskip
\noindent As we have already seen in paragraph 15, expressions (31)
$$
d \, \delta x_i \, + \, \sum\limits_{j,l=1}^n\Bigl\{ ^{jl}_{\, i} \Bigr\} \, d x_j  \, \delta x_l   = 0
$$
\vfil
\eject
\noindent for the second order differentials are no different from those we obtained by making Riemann's comprehensive definition explicit.
\par
Likewise, this paragraph also shows that from these expressions of the second order differentials we identically get (Ricci's lemma)
$$
\delta \, d s^2\, = \, d \, \delta s^2 =d\, \Phi = \delta \, \Phi = 0 \, ,
\leqno(48)
$$
\noindent where
$$
\Phi = 
 \sum\limits_{i,k=1}^n a_{ik}\, dx_i \, \delta x_k \, ,
$$
\noindent and $d s^2$, $\delta s^2$ obviously correspond to
$$
 \sum\limits_{i,k=1}^n a_{ik}\, dx_i \, d x_k  \qquad \hbox{{\rm and}} \qquad  \sum\limits_{i,k=1}^n a_{ik}\, \delta x_i \, \delta x_k \, .
$$
\par
As things stand, the meaning that has to be given to the trinomial that is considered by Riemann:
$$
R = \delta^2 \, d s^2 \, - \, 2 \, d\, \delta \Phi \, + \, d^2 \, \delta s^2 \, ,
$$
\noindent does not seem to be ambiguous. Thanks to equation (48) such meaning must imply that $R=0$.
\smallskip
\par
Instead Riemann states \footnote{($^{25}$)}{\foot{loc.~cit. ($^2$),~}\foot{p. 381.~}} that: ``Haec expressio (i.e., $R$) invenietur $=J$'' [where $J$ corresponds to equation (45)]. In his clarifications, Weber dwells on how the second differentials
should be introduced \footnote{($^{26}$)}{\foot{Adding, with no further justification, the additional conditions
$$
d^2 \, \delta s^2 = \delta^2 \, d s^2 = - \, 2 \, d\, \delta \Phi\, .
$$
\noindent Thanks to equation (48) (and provided that the formulae are read as they are actually written) everything vanishes.
}}, but, once he has given their explicit expression, he just says \footnote{($^{27}$)}{\foot{loc.~cit. ($^2$),~}\foot{p. 388.~}}: ``woraus man leicht den Ausdruck erh{\" a}lt $R=J$''.
\par
Probably, in Riemann's expression for $R$ there is just some writing flaw that clouds the idea that is behind it. I pride myself on having essentially reconstructed this idea but I wasn't able to correct the symbol. If this is possible, complete justice should be done to Riemann's genius also in regard to this specific point.
\par
I shall end with a remark about calculating the curvature regarding special variables, which is suggested by Riemann \footnote{($^{28}$)}{\foot{loc.~cit. ($^2$),~}\foot{p. 261.~}} and developed by Weber \footnote{($^{29}$)}{\foot{loc.~cit. ($^2$),~}\foot{p. 384-387.~}}. In the mean time, here is what it's all about.
\par
Let's choose coordinates $x_1, x_2, \dots , x_n$ so that, {\it at a particular point\/} $P$ all the $\Bigl\{ ^{jl}_{\, i} \Bigr\}$ symbols shall vanish (which is always possible, as Weber has clearly highlighted). Let's consider two $d x_i$, $\delta x_i$ independent differential systems and also
\vfil
\eject
\noindent that all second order differentials  $d^2 x_i$, $d\,\delta x_i$, $\delta\, d x_i$, $\delta^2 x_i$ shall vanish. Let's call the points of coordinates $x_i + dx_i$, $x_i + \delta x_i$ as $P'$ and $Q$; and the coefficients of the line element's square in $P'$ as $a'_{hk}$. Specifically, if we set
$$
(\delta s^2)_{P'} = \sum\limits_{h,k=1}^n a'_{hk} \, \delta x_h \, \delta x_k \, ,
$$
\noindent and apply the Taylor expansion to the $a'$ coefficients with respect to the $d$ increments, up to and including the second order. With such approximation we get
$$
(\delta s^2)_{P'} = \delta s^2 +{1\over 2}\, \sum\limits_{h,k,j,l=1}^n {{\partial^2a_{hk}}\over {\partial x_j \partial x_l}} \, dx_j\, dx_l\, \delta x_h \, \delta x_k \, ,
$$
\noindent where, obviously, $\delta s^2$ and the second derivatives refer to point $P$. For how the variables have been
set (as shown by Weber), there are some special relations among the values in point $P$ of the $a_{hk}$ coefficients' second derivatives. Bearing this in mind and with some transformation, we get
$$
(\delta s^2)_{P'} = \delta s^2 +{1\over 3}\, \sum\limits_{h,k,j,l=1}^n {1\over 2}\, \biggl[{{\partial^2a_{hk}}\over {\partial x_j \partial x_l}} + {{\partial^2a_{jl}}\over {\partial x_h \partial x_k}} - {{\partial^2a_{hj}}\over {\partial x_k \partial x_l}} - {{\partial^2a_{kl}}\over {\partial x_h \partial x_j}}\biggl] \, dx_j\, dx_l\, \delta x_h \, \delta x_k \, .
$$
\par
As mentioned earlier, when the $x$ particularized variables are used, the summation can be considered as the expression (according to formula (45)) for $- I$.
\par
Thus, considering equation (47), we get equation
$$
{{(\delta s^2)_{P'} - \delta s^2}\over {(ds\, \delta s \,\sin \psi)^2}} = - {1\over 3}\, K\, ,
\leqno(49)
$$
\noindent that Riemann describes in words in the mentioned passage (by multiplying both sides of the equation by $4$ so that the area of triangle $PP'Q$ shall stand out in the denominator).
\smallskip
\par
Finally I shall come to my finding:
\par
If we call $Q^{\ast}$ the extremity of line element $(\delta s)_{P'}$ (that corresponds to increments $\delta x_i$) then equation (49) can be written as
$$
{{  \overline{P'Q^{\ast}}^2 - \overline{PQ}^2}\over {(ds\, \delta s \,\sin \psi)^2}} = - {1\over 3}\, K\, ;
\leqno(49')
$$
\noindent while equation (46) (having changed its sign) provides equation
$$
{{  \overline{P'Q'}^2 - \overline{PQ}^2}\over {(ds\, \delta s \,\sin \psi)^2}} = - K\, .
\leqno(46')
$$
\noindent As it can be seen, the ratio of the these equations' right-hand sides is $1$ to $3$. The fact they don't coincide is clearly because point $Q'$ (that is obtained with the invariant procedure and corresponds to the parallelogramoid's fourth vertex) is distinct from Riemann's point $Q^{\ast}$ that is analytically defined with reference to special variables.
\smallskip
\par
In order to pinpoint the difference between the formulae, also our procedure will benefit from being carried out with Riemann's special variables (of course thanks to its invariant nature).
\vfil
\eject
\noindent As equations (31) refer to the point $P$, they provide
$$
d^2 x_i =d\delta x_i = \delta dx_i = \delta^2 x_i =0\, ;
$$
{\it but this does not mean that also their higher order differentials\/} (i.e., $\delta d^2 x_i$, $d^2\delta x_i$, etc.) must also vanish in the same point. Instead, Riemann's calculation stands on the hypothesis that all differentials above the first order shall vanish: this hypothesis is also legitimate but it does not have a invariant nature (when the variables are changed). It shouldn't then surprise if the results are not the same. While instead the accidental analogy between formulae (49′) and (46′) (where the righthand sides only differ by a numerical factor) should be noted.
\par\bigskip Padua, November 1916.
\par\medskip \rightline{Tullio Levi-Civita}

\vbox to 9.0 cm {}
\centerline{$\underline {\qquad\qquad\qquad\qquad}$}

\end


Let $P'$ and $Q$ denote the points of coordinates $x_i + dx_i$, $x_i + \delta x_i$; and $a'_{hk}$ the coefficients of the squared line element in $P'$. In particular, setting
$$
(\delta s^2)_{P'} = \sum\limits_{h,k=1}^n a'_{hk} \, \delta x_h \, \delta x_k \, ,
$$
\noindent let us apply to $a'$ the the Taylor expansion with respect to the increments $d$, up to the second order.  In such approximation one has
$$
(\delta s^2)_{P'} = \delta s^2 +{1\over 2}\, \sum\limits_{h,k,j,l=1}^n {{\partial^2a_{hk}}\over {\partial x_j \partial x_l}} \, dx_j\, dx_l\, \delta x_h \, \delta x_k \, ,
$$
\noindent $\delta s^2$ and the second derivatives referring, of course, to $P$. As shown by Weber, due to the way the variables were fixed, special relations hold between the values of the second derivatives of the $a_{hk}$ in $P$. Taking them into account, one finds, with some manipulation, 
$$
(\delta s^2)_{P'} = \delta s^2 +{1\over 3}\, \sum\limits_{h,k,j,l=1}^n {1\over 2}\, \biggl[{{\partial^2a_{hk}}\over {\partial x_j \partial x_l}} + {{\partial^2a_{jl}}\over {\partial x_h \partial x_k}} - {{\partial^2a_{hj}}\over {\partial x_k \partial x_l}} - {{\partial^2a_{kl}}\over {\partial x_h \partial x_j}}\biggl] \, dx_j\, dx_l\, \delta x_h \, \delta x_k \, .
$$
\par The sum can be looked at as the expression which, as the basis of formula (45), is taken on by when variables specified as above are adopted.
\par Therefore, taking into account (47), we get
$$
{{(\delta s^2)_{P'} - \delta s^2}\over {(ds\, \delta s \,\sin \psi)^2}} = - {1\over 3}\, K\, ,
\leqno(49)
$$
\noindent which Riemann, in the quoted passage, states in words (multiplying both sides by $4$, in order to show up the area of the triangle $PP'Q$ in the denominator).
\smallskip
\par I come, at last, to my point:
\par If $Q^{\ast}$ denotes the extremum of the line element $(\delta s)_{P'}$ (corresponding to the increments $\delta x_i$), eq. (49) can be written
$$
{{  \overline{P'Q^{\ast}}^2 - \overline{PQ}^2}\over {(ds\, \delta s \,\sin \psi)^2}} = - {1\over 3}\, K\, ;
\leqno(49')
$$
\noindent whereas eq. (46) (with an overall change of sign) reads
$$
{{  \overline{P'Q'}^2 - \overline{PQ}^2}\over {(ds\, \delta s \,\sin \psi)^2}} = - K\, .
\leqno(46')
$$
\noindent As can be seen, the right-hand sides are in the ratio $1$ to $3$. The lack of coincidence is manifestly due to the fact that the point $Q'$ (fourth vertex of the parallelogrammoid), which is reached through the invariant procedure, is well distinct from Rieman's point $Q^{\ast}$ analytically defined with reference to particular variables.
\par To localize the discrepancy about the formulae, it helps to work out our procedure too (as is of course allowed given its invariant character) in Riemann's special variables.

\vfil
\eject
\noindent Eqs. (31) give then, in so far as they refer to the point $P$,
$$
d^2 x_i =d\delta x_i = \delta dx_i = \delta^2 x_i =0\, ;
$$
{\it but it does not follow that the higher differentials}, such as $\delta d^2 x_i$, $d^2\delta x_i$, etc. must vanish at the same point as well. Riemann's calculation on the contrary is based on the hypothesis that all differentials of an order higher than the first must vanish: a legitimate hypothesis too, but not one endowed with an invariant character (with respect to changes of variables). Therefore, it should not come as a surprise that the results are different: one should rather notice the fortuitous analogy between formulae $(49')$ and $(46')$, whose right-hand sides differ only by a numerical factor.

\par\bigskip Padova, November 1916.
\par\medskip \rightline{Tullio Levi-Civita}

\vbox to 9.0 cm {}
\centerline{$\underline {\qquad\qquad\qquad\qquad}$}

\end

\vfil
\eject
\centerline{Translators' notes}
\bigskip
\noindent [1] In the original text, the index $_h$ is missing from the expression $\delta x_h$ in eq. (32).
\par\medskip
\noindent [2] mmm

\end